\documentclass[12pt]{article}

\usepackage{epsfig,amsfonts,newlfont,rotate,float,amssymb}

\DeclareMathAlphabet{\mathitbf}{T1}{cmr}{bx}{it}

\newcommand{\e}{\mathrm e}
\newcommand{\openone}{{\mbox{\boldmath $1$}}}
\begin{document}

\title{
Phase-diagram and quasiparticles of a lattice SU(2)
scalar-fermion model in (2+1) dimensions.
}
 
\author{
J.L.~Alonso$^1$, Ph.\ Boucaud$^2$, \\
V.~Mart\'{\i}n-Mayor$^3$ and A.J.~van der Sijs$^4$
}

\date{\today}
\maketitle
\thispagestyle{empty}

\noindent {\small {\it $^1$Departamento de F\'\i sica Te\'orica, Universidad de Zaragoza,
50009 Zaragoza, Spain.}} 

\noindent {\small {\it $^2$LPTHE, Universit\'e de Paris XI,
91405 Orsay Cedex, France.}}

\noindent {\small {\it $^3$Dipartimento di Fisica and Infn, Universit\`a di Roma, La Sapienza P. A. Moro 2, 00185 Roma.}}

\noindent {\small {\it $^4$Swiss Center for Scientific Computing, ETH-Z\"urich,
ETH-Zentrum, CH-8092 Z\"urich, Switzerland.}

\noindent {\small {\bf e-mail:} {\it 
$^1$buj@gteorico.unizar.es, $^2$phi@qcd.th.u-psud.fr,$^3$Victor.Martin@roma1.infn.it,}}

\noindent {\small {\it $^4$arjan@scsc.ethz.ch}}

\begin{abstract}

The phase diagram at zero temperature of a lattice SU(2) scalar-fermion model 
in (2+1) dimensions is studied numerically and with Mean-Field methods. 
Special attention is devoted to the strong coupling regime. We have developed 
a new method to adaptate the Hybrid Monte Carlo algorithm to the O(3) 
non-linear $\sigma$ model constraint.
The charged excitations in the various phases are studied at the Mean-Field level. 
Bound states of two charged fermions are found in a strongly coupled 
{\it paramagnetic\/} phase. On the other hand, in the strongly coupled 
antiferromagnetic phase fermionic excitations around momenta 
$(\pm\frac{\pi}{2},\pm\frac{\pi}{2},\pm\frac{\pi}{2})$ emerge. 

\end{abstract}
  
\vskip 5 mm

\medskip


\vskip 5 mm

\newpage


\section{Introduction}
\label{intro}

The model we are going to study was proposed in references \cite{LETTER}
and~\cite{HEPLAT} as a natural extension of the lattice O(3) non-linear 
$\sigma$-model in (2+1)~dimensions to include charge carriers. It is a 
lattice model of interacting spins and Dirac fermions in (2+1)~dimensions, 
with only two free parameters in addition to the temperature: a 
nearest-neighbour spin coupling and a spin-fermion coupling. The model 
describes quantitatively some of the features of the doped copper oxide 
compounds~\cite{LETTER,HEPLAT}.

In the present article we want to present a careful, detailed
discussion of the model, its symmetries, and its properties, and give
full technical details and results of the Mean-Field (MF) and Monte Carlo (MC) 
calculations, some of which were reported in Ref.\ \cite{LETTER}.
In this paper, our Mean-Field and numerical studies will be limited to the 
zero-temperature case, corresponding to infinite Euclidean time direction.

The remainder is laid out as follows. In Section~\ref{TM} we present our model, 
discuss the choice of lattice fermions, comment on the symmetries of the model, 
give its phase diagram and prove the reality of the fermion determinant, 
even in the presence of a chemical potential.  
In Section~\ref{MFPD} we examine the phase diagram of the model in the MF 
approximation. Our MF calculations are based on small- and large-$y$ 
expansions combined with saddle points methods. The method allows us to handle
(products of) fermionic variables occurring in the expansion of the fermion
determinant in a well-defined way. In Section~\ref{MCPD} we use MC simulations
to complete the study of the phase diagram.  For this purpose we have developed 
a new method that exactly solves the technical problem related to the length-1 
constraint on the spin variables \cite{PARISI}.  Section~\ref{sectMFexc} is 
devoted to a study of the relevant excitations in the different phases of the 
system, at the MF level. A crucial result is the dynamical generation of spin 
singlet bosonic bound states of charged fermions in the so-called paramagnetic 
strong (PMS) phase. At the MF level we have not found evidence for {\em fermionic\/} 
excitations at {\it zero} temperature in this PMS phase.  Another interesting 
result is the emergence of fermionic excitations around momenta
$(\pm\frac{\pi}{2},\pm\frac{\pi}{2},\pm\frac{\pi}{2})$ in the strongly coupled 
Antiferromagnetic (AFM) phase \cite{CHUBUKOV}. 
Finally, Section~\ref{CONC} is devoted to our conclusions and projects.

\section{The model: Formulation, Symmetries, Phase Diagram}
\label{TM}

The  model is defined by the following (2+1)-dimensional lattice 
Euclidean (imaginary time) path integral,

\begin{equation}
Z = \int\!D{\mbox{\boldmath $\phi$}} \, D\bar\psi \, D\psi \, \exp(-S)
\label{Z}
\end{equation}
with action,
\begin{equation}
S = -\sum_{x,\mu} k \, {\mbox{\boldmath $\phi$}}_x\cdot {\mbox{\boldmath $\phi$}}
_{x+\hat\mu}+
 \sum_{x,\mu} \frac{\rho}{2} \bar\psi_x \gamma^\mu (\psi_{x+\hat\mu} -
    \psi_{x-\hat\mu}) 
+\sum_{x} \lambda \, \bar\psi_x {\mbox{\boldmath $\phi$}}_x \cdot {\mbox{\boldmath $
\tau$}} \psi_x\label{ACCIONZ}
    \, .
\end{equation}
We use this expression as our starting point, but it should be noted
that the model
{\it
depends only on the ratio $y=\lambda/\rho$, through a change in
the normalization of the fermion field}. In terms of the effective
spin-fermion coupling $y$, we get:

\begin{equation}
\label{ACCION}
S=-\sum_{x,\mu} k \, {\mbox{\boldmath $\phi$}}_x\cdot {\mbox{\boldmath $\phi$}}
_{x+\hat\mu}+
 \sum_{x,\mu} \, \frac{1}{2}\, \bar\psi_x \gamma^\mu (\psi_{x+\hat\mu} -
    \psi_{x-\hat\mu})
+\sum_{x} \, y\,  \, \bar\psi_x {\mbox{\boldmath $\phi$}}_x \cdot {\mbox{\boldmath $
\tau$}} \psi_x
    \, .
\end{equation}

Here $x$ runs over a (2 + 1)-dimensional cubic Euclidean space-time lattice.
Each $\psi_x$ is a fermionic four-spinor as a shorthand for two flavours of 
two-component Dirac spinors. Both flavours are taken in the same irreducible spinor
representation, with $2\times 2$ gamma matrices taken as the Pauli matrices 
$\sigma^\mu$. The $4\times 4$ matrices $\gamma^\mu$ in Eq.\ (\ref{ACCION}) 
have the form

\begin{equation}
\gamma^\mu = \left( \begin{array}{cc} \sigma^\mu & 0 \\ 0 & \sigma^\mu
    \end{array} \right) \ \ \ \ \ \ \ \ \ \ \ \mu = 1,2,3.
\label{gammas}
\end{equation}

The kinetic term for the fermions is of the nearest-neighbour (hopping) form. 
Lattice fermions defined in this way undergo species doubling in the 
perturbative continuum limit~\cite{CONTINUUM}. For two reasons we are going to 
leave this matter aside in this work. First, we are particularly interested in the 
strong coupling non-perturbative regime where more of the interesting physics 
occurs~\cite{LETTER, HEPLAT}. In this strong coupling regime all the fermions, 
the physical one as well as the doublers, decouple in the continuum limit~\cite{DE}. 
Second, this model described qualitatively some of the features of the doped 
copper oxide compounds~\cite{LETTER, HEPLAT}, where the lattice space is given 
by the material.

The three-component ${\mbox{\boldmath $\phi$}}$ are real scalar bosonic variables, 
subject to the constraint ${\mbox{\boldmath $\phi$}}^2=1$, as in the O(3) non-linear
$\sigma$-model. The last term in Eq. (\ref{ACCION}) describes the interaction 
between ${\mbox{\boldmath $\phi$}}$ and the Dirac fermions, which is diagonal 
in Dirac space. The Pauli matrices $\tau^a$ act in flavour space.

Let us now consider the symmetries of (\ref{ACCION}).
First of all, we have the usual cubic symmetry.
Next, there is a discrete parity symmetry, which in (2+1) dimensions
is defined as the reflection of one of the spatial axes, say the $x$-axis.
Under this parity symmetry, the fermions can be seen to transform as
\begin{equation}
\psi \rightarrow \sigma_1 \, \psi, \ \ \ \ \ \ \ \ 
\bar\psi \rightarrow - \bar\psi \, \sigma_1
 \, , \label{psipar}
\end{equation}
so ${\mbox{\boldmath $\phi$}}$ is a pseudoscalar in this sense.
In addition, the action (\ref{ACCION}) is invariant under an SU(2) flavour
symmetry in which $\psi$ transforms as the fundamental representation and
${\mbox{\boldmath $\phi$}}$ transforms as the adjoint one.
Note that by requiring the two fermion flavours to have the same Lorentz
structure (that is, by choosing the $\gamma$'s as in (\ref{gammas}))
no fermion mass term is allowed which preserves the above
symmetries~\cite{APPELQUIST}.

There are two more discrete symmetries of our model (\ref{ACCION}),
which will be useful in the MF calculation of the phase-diagram.
The first one is
\begin{equation}
Z(k,y) = Z(k,-y)
 , \label{symmZ1}
\end{equation}
which becomes clear if we make the change of variables
\begin{eqnarray}
\label{symmfields1}
\psi_x \rightarrow \exp\left(i\frac\pi2 \epsilon_x\right) \, \psi_x, \ \ \ \ 
\bar\psi_x \rightarrow \exp\left(i\frac\pi2 \epsilon_x\right) \, \bar\psi_x,
\end{eqnarray}
where
\begin{equation}
\epsilon_n = (-1)^{\sum_\mu x_\mu}
 . \label{epsilonn}
\end{equation}
This implies that $Z(k,y)$ is a function of $y^2$ only and we can restrict
ourselves to $y>0$.

In addition, there is a symmetry
\begin{equation}
Z(k,y) = Z(-k,-iy)
 , 
\label{symmZ2}
\end{equation}
as can be seen by making the substitutions
\begin{eqnarray}
\label{symmfields2}
\psi_x \rightarrow \exp\left(i\frac\pi4 \epsilon_x\right) \, \psi_x, \ \ \ \ 
\bar\psi_x \rightarrow \exp\left(i\frac\pi4 \epsilon_x\right) \, \bar\psi_x, \ \ \ \ 
{\mbox{\boldmath$\phi$}}_x \rightarrow \epsilon_x \, {\mbox{\boldmath$\phi$}}_x
 . 
\label{SIMSTAGG}
\end{eqnarray}

The latter symmetry implies that the lattice regularization of
the non-linear $\sigma$-model, $y$=0 (or $y$=$\infty$, see Sections
\ref{MFPD}, \ref{MCPD}), is equally valid in a ferromagnetic 
or an antiferromagnetic phase.

In order to perform computations in models of this type,
one has to integrate out the fermions. This integration leads to a 
${\mbox{\boldmath $\phi$}}$-dependent fermion determinant.
It is important to know whether this determinant is a real number.
To study this, let us write down the original fermion matrix 
(Latin letters $x, y, \ldots$ will refer to lattice points, 
$i, j, \ldots$, will represent flavour indices, while Greek
letters $\alpha, \beta, \ldots$ are used for Dirac indices):

\begin{eqnarray}
\hat M_{x\alpha i;y\beta j}&=&K_{x\alpha i;y\beta j}+Y_{x\alpha
  i;y\beta j},
  \label{matrixM} \\
K_{x\alpha i;y\beta j}&=&\frac{1}{2}\sum_{\mu}\left(\delta_{x+\mu,y}-
\delta_{x-\mu,y}\right)\sigma^{\mu}_{\alpha\beta}\,\delta_{ij},
  \label{MATRIZK} \\
Y_{x\alpha i;y\beta j}&=&y\,\delta_{xy}\left({\mbox{\boldmath $\phi$}}\cdot
{\mbox{\boldmath {$\tau$}}}\right)_{ij}\,\delta_{\alpha\beta}.
  \label{matrixY}
\end{eqnarray}
Keeping in mind that for Pauli matrices $\sigma_2\sigma_i\sigma_2=-\sigma_i^*$,
where $^*$ means complex conjugation, and that 
$\left[{\mbox{\boldmath $\gamma$}},{\mbox{\boldmath $\tau$}}\right]=0$,
one easily proves that, for real $y$,
\begin{equation}
\sigma_2\tau_2\left(K\, +\, Y\right)\sigma_2\tau_2=
-\left(K^*\, +\, Y^*\right).
\label{REALDET}
\end{equation}
Therefore,
\begin{equation}
\det\left(K\, +\, Y\right)=\det\left(-K^*\, -\, Y^*\right)=
\left[\det\left(K\, +\, Y\right)\right]^*,
\end{equation}
{\em i.e.} the determinant is real.
Thus, by doubling the number of fermion families, we obtain a positive
determinant. Had we introduced a chemical potential, $\mu$, the only change
would be the introduction of $\e^{\pm\mu}$ on the temporal links of
the kinetic matrix~\cite{KARSCH}. The essential requirement
for Eq.\ (\ref{REALDET}) to hold (that the only non-real numbers are in 
{\mbox{\boldmath $\gamma$}},{\mbox{\boldmath $\tau$}}) is thus
not endangered by the chemical potential and the determinant is still real. 

\begin{figure}[htb]
\begin{center}
\leavevmode
\centering\epsfig{file=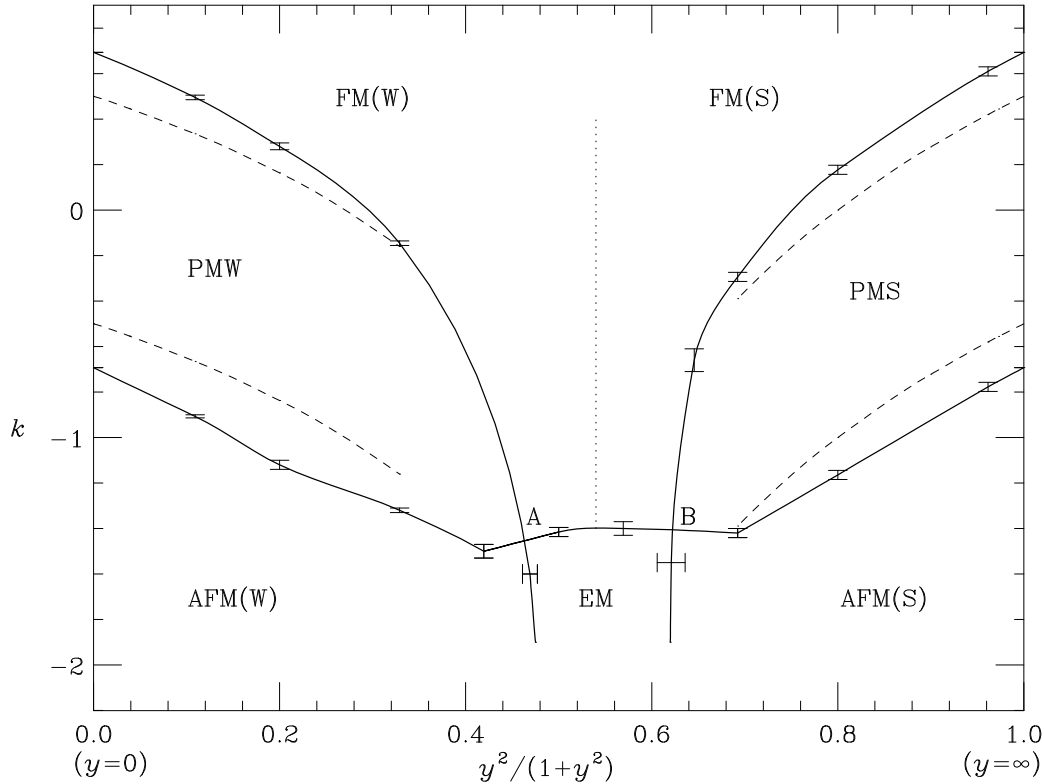,width=0.7\linewidth,angle=90}
\end{center}
\caption{Phase diagram of the action (\protect\ref{ACCION}), for two
fermion families.
Dashed lines are from the MF calculation,
solid lines from a MC calculation on an $8^3$ lattice.}
\label{PHASES}
\end{figure}

The phase diagram of the model at zero temperature is shown 
in fig.\ \ref{PHASES}. Notice that it is very
similar to the phase diagram of (chiral) Yukawa models for the electroweak
sector of the Standard Model of elementary particle 
interactions~\cite{shigemitsubocketal}.
At $y=\infty$ and at $y=0$ we
recover the non-linear $\sigma$-model (see sections \ref{MFPD},\ref{MCPD}) with its well known
paramagnetic (PM), ferromagnetic (FM) and antiferromagnetic (AFM) phases. 
At finite $y$, we expect these phases to extend into the ($k$,$y$) plane.
One of its most important features is that there are two mutually
disconnected paramagnetic phases, one at weak coupling (called PMW)
and one at strong coupling (PMS).
One sees that the PMW-FM and the PMW-AFM transition lines meet in a point
{\bf A}, where this disordered phase ends.
In the strong coupling
sector of the phase diagram, a similar behaviour is found, with the
two transition lines meeting at point {\bf B}.
This observation means that one may expect totally different behaviour
in each of the two paramagnetic phases.
This is indeed the case, as we shall see later.

As there is no evidence for a phase transition between the strong- and
weak-coupling regions of the FM and AFM phases,
we name them FM(W) and FM(S), AFM(W) and AFM(S) (note the parentheses).
There may be crossovers between these regions, though.

Between the points {\bf A} and {\bf B}, we find a phase where both
the magnetization and the staggered magnetization are different
from zero. We name this phase {\it exotic magnetic} (EM).
An appealing possibility is that it corresponds to a helicoidal phase.
We expect the EM phase to disappear for large enough $-k$, but we have not
explored this numerically.

\section{Mean Field Calculations of the Phase Diagram}

\label{MFPD}

Our aim in this section is to determine the zero-temperature
phase diagram of the model in the $y$-$k$ plane (cf.\ Fig.\
\ref{PHASES}), using Mean-Field techniques.
These calculations already provide a lot of insight, especially
for the strong coupling region.  They will be contrasted with
numerical simulations for the phase diagram in Sect.\ \ref{MCPD},
and they will be extended to a study of the relevant charged
(quasi-particle) excitations in Sect.\ \ref{sectMFexc}.

Our Mean-Field calculations are based on small- and large-$y$
expansions combined with the saddle point methods described in Ref.\
\cite{drouffezuber}.  This approach guarantees a systematic expansion
in $1/d$, which is particularly important for operators which are zero
to lowest-order.  Our particular method furthermore allows us to
handle (products of) fermionic variables occurring in the expansion of
the fermion determinant in a well-defined way.  These techniques have
been developed and applied in the context of similar lattice models
\cite{plzar,class} of the Electroweak sector of the Standard Model of
elementary particle interactions, and in the study of the
antiferromagnetic $\phi^4$ model~\cite{POLONYI}.

We shall first concentrate on the small-$y$ region, and
incorporate the fermion determinant up to ${\cal O}(y^2)$.

In order to apply the saddle-point method, the integration over the
fields must be unrestricted.
We therefore need to replace the integration over the spin
vectors ${\mbox{\boldmath $\phi$}}$, constrained by the condition $|{\mbox{\boldmath $\phi$}}| = 1$, with an
integration over unconstrained variables ${\mbox{\boldmath $\xi$}}$.
This is done by multiplying the functional integrand in Eq.\ (\ref{Z}) by
\begin{eqnarray*}
1 &=& \int\! D\xi \, \delta({\mbox{\boldmath $\phi$}} - {\mbox{\boldmath $\xi$}})
\equiv \prod_n \prod_{a=1}^3 \int_{-\infty}^\infty \! d\xi_x^a
\, \delta(\phi_x^a - \xi_x^a)
\\ 
 &=&
\prod_x \prod_{a=1}^3 \int_{-\infty}^\infty \! d\xi_x^a
 \, \int_{-\infty}^\infty \! \frac{dA_x^a}{2\pi}
 \, \exp[iA_x^a (\phi_x^a - \xi_x^a)]\ ,
\end{eqnarray*}
and replacing a conveniently chosen subset of the ${\mbox{\boldmath $\phi$}}$ variables
in the action $S$ with ${\mbox{\boldmath $\xi$}}$ fields.
We obtain
\begin{eqnarray}
\label{Z2}
Z &=&
\int \! \frac{D\xi DA}{(2\pi)^3}
 \, \exp\left[ k \sum_{x,\mu} {\mbox{\boldmath $\xi$}}_x \cdot {\mbox{\boldmath $\xi$}}_{x+\mu}
 - i \sum_x {\mbox{\boldmath $A$}}_x \cdot {\mbox{\boldmath $\xi$}}_x \right]
 \nonumber \\
 &\times &
\int \! D\bar\psi D\psi
 \, \exp\left[ -\sum_{x,\mu} \frac12 \bar\psi_x \gamma^\mu (\psi_{x+\hat\mu} -
    \psi_{x-\hat\mu}) \right]\\\nonumber
 &\times & \prod_x \left\{ \int \! \frac{d{\mbox{\boldmath $\phi$}}_x}{4\pi}
 \, \exp\left[ i{\mbox{\boldmath $A$}}_x \cdot {\mbox{\boldmath $\phi$}}_x
        - y \, \bar\psi_x \, {\mbox{\boldmath $\phi$}}_x\cdot\vec\tau \, \psi_x \right] \right\}
 . 
\end{eqnarray}
Note that both the ${\mbox{\boldmath $\xi$}}$ fields and the auxiliary fields ${\mbox{\boldmath $A$}}$ are
unconstrained.

Now we have to integrate out the constrained variables
$\phi^a_n$ (as well as the fermions), before the mean fields can be introduced.
Let us concentrate on a single ${\mbox{\boldmath $\phi$}}_n$ integration in Eq.\ (\ref{Z2}),
dropping the subscripts $n$ for simplicity.
First, we perform an expansion in powers of $y$.
We can write
\begin{eqnarray}
&&\int \! \frac{d{\mbox{\boldmath $\phi$}}}{4\pi}
 \, \exp\left[ i{\mbox{\boldmath $A$}} \cdot {\mbox{\boldmath $\phi$}}
        - y \, \bar\psi \, {\mbox{\boldmath $\phi$}}\cdot\vec\tau \, \psi \right]
 \nonumber \\
&&=
\exp\left[ u(i{\mbox{\boldmath $A$}}) \right]
 \, \exp\left[ -y\, Q^a \cdot \langle\phi^a\rangle_{i{\mbox{\boldmath $A$}}}
  + \frac12 y^2 \, Q^a Q^b T^{ab} + {\cal O}(y^3) \right]
  , \label{zphi2}
\end{eqnarray}
where we have defined
$$Q^a = \bar\psi \tau^a \psi\ \  ,\ \  
u(i{\mbox{\boldmath $A$}}) = \ln \, \int \! \frac{d{\mbox{\boldmath $\phi$}}}{4\pi}
\, \exp\left[ i{\mbox{\boldmath $A$}} \cdot {\mbox{\boldmath $\phi$}} \right]
\ ,\  T^{ab} =\langle \phi^a \phi^b\rangle_{i{\mbox{\boldmath $A$}}} - 
\langle\phi^a\rangle_{i{\mbox{\boldmath $A$}}}
\langle\phi^b\rangle_{i{\mbox{\boldmath $A$}}},$$
and we have introduced the notation
$$
\langle{O}\rangle_{i{\mbox{\boldmath $A$}}} =
\left.{\int \! \frac{d{\mbox{\boldmath $\phi$}}}{4\pi}}
\, O \, \exp\left[ i{\mbox{\boldmath $A$}} \cdot {\mbox{\boldmath $\phi$}} \right] \right/
{\int \! \frac{d{\mbox{\boldmath $\phi$}}}{4\pi}
\, \exp\left[ i{\mbox{\boldmath $A$}} \cdot {\mbox{\boldmath $\phi$}} \right] }
\, .$$

In addition, we introduce a Hubbard-Stratonovich vector parameter
${\mbox{\boldmath $\lambda$}}$ to deal with the quartic fermion term in Eq.\ (\ref{zphi2}),
\begin{eqnarray}
\exp\left[ \frac12 y^2 \, \sum_{a,b} \, Q^a Q^b T^{ab}\right]
&=&
\int \! \frac{d{\mbox{\boldmath $\lambda$}}}{(2\pi)^{3/2}}
 \, \exp\left[ - \frac12 \sum_a \lambda^a \lambda^a
\right]\\\nonumber
&\times&\exp\left[y \sum_{ab} \left(\sqrt{T}\, \right)^{ab}
    Q^a \lambda^b\right]\, . 
\label{HSlambda}
\end{eqnarray}
(Note that the matrix  $T$ is self adjoint, and positive definite if 
${\mbox{\boldmath $A$}}$ is imaginary, so the square root is well defined). 
Thus, up to this order in $y^2$,
the action is bilinear in the fermion fields.

Carrying out the fermion integration in Eq.\ (\ref{Z2}) now gives det $M$,
where
\begin{equation}
M_{x,\alpha,i;y,\beta,j} = K_{x,\alpha,i;y,\beta,j}
  + y \delta_{xy} \delta_{\alpha\beta} \sum_a
 (\langle \phi^a_x \rangle_{iA_x} - 
\sum_b\left(\sqrt{T_x}\right)^{ab} \lambda^b) 
\tau^a_{ij}
 \, . \label{Mdef}
\end{equation}
The matrix $K$ has been defined in Eq. (\ref{MATRIZK}).

The mean fields are the field values at the saddle point of the free energy
\begin{equation}
-F = \sum_x u(i{\mbox{\boldmath $A$}}_x) + k \sum_{x,\mu} {\mbox{\boldmath $\xi$}}_x \cdot {\mbox{\boldmath $\xi$}}_{x+\mu}
 - i \sum_x {\mbox{\boldmath $A$}}_x \cdot {\mbox{\boldmath $\xi$}}_x
 - \frac12 \sum_x {\mbox{\boldmath $\lambda$}}_x^2 + \mathrm{Tr\,} \log M
 \, . \label{Fdef}
\end{equation}
A choice of the mean fields should be done at this point, as we cannot
calculate $\log M$ for general $\{{\mbox{\boldmath $A$}}_x\, ,\, 
{\mbox{\boldmath $\lambda$}}_x\}$. An appropriate choice for the study
of a PM-FM phase transition is

\begin{eqnarray}
{\mbox{\boldmath $A$}}_x &=& (0,0,-i\alpha)
 \, , \\\nonumber
{\mbox{\boldmath $\xi$}}_x &=& (0,0,v)
 \, ,  \\\nonumber
{\mbox{\boldmath $\lambda$}}_x &=& (0,0,\lambda)
 \, , 
\end{eqnarray}
in terms of which ($N$ is the lattice volume)
\begin{equation}
F/N = - u(\alpha) - kdv^2 + \alpha v + \frac12 \lambda^2
    - \frac1N\, \mathrm{Tr\,} \log M
 \, , \label{FoverN}
\end{equation}
with $\alpha$, $v$ and $\lambda$ satisfying
the saddle point equations
\begin{equation}
\left.\nabla F\, \right|_{(\alpha,v,\lambda)}=0
 \, . \label{sadd}
\end{equation}

The fermion matrix, $M(\alpha,v,\lambda)$, can be calculated
in momentum space, where it is diagonal in its momentum indices.
One easily finds 
\begin{equation}
\mathrm{det\,} M = \exp\left[ 2 \sum_p \log \frac{ \sum_{\mu=1}^3 \sin^2 p_\mu
  + y^2 \left(u'(\alpha) - \lambda \sqrt{u''(\alpha)}\right)^2 }
    {\sum_{\mu=1}^3 \sin^2 p_\mu }  \right]
 \, , \label{Mresult}
\end{equation}
where we have divided out the determinant for free fermions.
We need only the leading ${\cal O}(y^2)$ contribution to the exponent,
hence the mean field free energy becomes, in the infinite volume limit:
\begin{equation}
F/N = - u(\alpha) - kdv^2 + \alpha v + \frac12 \lambda^2
    - 2 y^2 \left(u'(\alpha) - \lambda \sqrt{u''(\alpha)}\right)^2 \, C_0
 \, , \label{FoverN2}
\end{equation}
where
\begin{equation}
C_0 \ =\ \int_{-\pi}^\pi \! \frac{d^3p}{(2\pi)^3}
 \, \frac1{\sum_{\mu=1}^3 \sin^2 p_\mu} \ =\ 1.0109240
 \, . \label{C0}
\end{equation}

Next, we shall discuss the actual solutions to Eqs.\ (\ref{sadd}).
From
$u(\alpha) \ =\ \ln \, (\sinh\alpha / \alpha)$, one easily 
finds that $\alpha = v = \lambda = 0$ always fulfill them.
For small $k$, $y$, it is a true minimum
of the free energy.
This characterizes a paramagnetic (PM) phase, since none of the fields
develops an expectation value.

For larger values of $k$ and $y$, there is another, non-trivial solution,
corresponding to a ferromagnetic (FM) phase.
It emerges when a negative mode in $F/N$ starts to develop, as a function
of the mean fields, and the
transition between the two regions is given by the condition ($F''$ is
the Hessian matrix)
\begin{equation}
\left.\mathrm{det\,}F''\right|_{(\alpha=0,v=0,\lambda=0)} = 0
 \, . \label{transcond}
\end{equation}
This condition is satisfied for $F/N$ of Eq.\ (\ref{FoverN2})
if 
\begin{equation}
k = \frac3{2d} - \frac{2\, C_0}d y^2
 \, . \label{ky2}
\end{equation}

This curve defines the phase transition line between the PM and FM
phases in the small-$y$ region.
Using the symmetry (\ref{symmZ2}), we deduce that there is a similar
transition separating the PM and AM phases,
\begin{equation}
k = - \frac3{2d} - \frac{2\,C_0}d y^2
 \, . \label{ky2a}
\end{equation}
Let us finally remark that in the presence of $N_f$ such fermion fields
we would have $N_f$ factors of det~$M$, leading to a multiplication of
$C_0$ by $N_f$ in Eqs.\ (\ref{ky2}) and (\ref{ky2a}).

The large-$y$ region is easier to deal with.
Here it is convenient to integrate out the fermions directly
in Eq.\ (\ref{Z2}), leading to (summation over repeated index is carried)
\begin{eqnarray}
\mathrm{det\,} M_{x,\alpha,i;y,\beta,j} &=&
\mathrm{det\,} \left(K_{x,\alpha,i;y,\beta,j}
+ y \delta_{\alpha\beta} \delta_{xy} \sum_a \phi^a_x \tau^a_{ij} \right)
 \label{Mylarge0} \\
&=&
\mathrm{det\,} \left(y\delta_{\alpha\gamma} \delta_{xz} \sum_a \phi^a_x
  \tau^a_{ik} \right) \nonumber \\
&&\times\
\mathrm{det\,} \left( \delta_{zy} \delta_{\gamma\beta} \delta_{kj}
 + \frac1y \sum_b \phi^b_x \tau^b_{kl} K_{z,\gamma,l;y,\beta,j}  \right)
 \, . \label{Mylarge}
\end{eqnarray}
Here we have used that $(\sum_a \phi^a \tau^a)^2 = \openone$ (recall
the ${\mbox{\boldmath $\phi$}}$'s are unit vectors).
Now we can expand log(det $M$) in powers of $1/y$.
The ${\cal O}(1/y)$ term vanishes by virtue of $K_{xx} = 0$.
To second order one obtains
\begin{eqnarray}
\log \mathrm{det\,} M &=& \log y^{4N} + \mathrm{Tr\,} \left(
 -\frac1{2y^2} \sum_a \phi^a_x \tau^a_{ki} K_{x\alpha i;t\gamma l}
\sum_b \phi^b_t \tau^b_{lj} K_{t\gamma j;y\beta p}\right)
  \label{trlogM0} \\
&=&
\log y^{4N} + \frac1{y^2} \sum_{x,\mu}
    {\mbox{\boldmath $\phi$}}_x \cdot {\mbox{\boldmath $\phi$}}_{x+\hat\mu}
 \, . \label{trlogM2}
\end{eqnarray}
Here, $\log y^{4N}$ is an irrelevant constant that can be
dropped. Notice also that
this expression will acquire a prefactor $N_f$ if there are $N_f$ identical
fermion flavours.
One sees that, up to ${\cal O}(1/y^2)$, the only effect of the fermion
determinant is a renormalization of the scalar hopping parameter of the
O(3) model,
\begin{equation}
k \rightarrow k + N_f \, \frac1{y^2}
  \, . \label{kRenorm}
\end{equation}

Note that we did not introduce any mean fields to derive this result.
The usual MF treatment of the O(3) model with this renormalized coupling
now immediately gives us the required phase transition lines in the large-$y$
region of our model:
\begin{equation}
k = \pm \frac3{2d} - N_f \, \frac1{y^2}
  \, . \label{ky2b}
\end{equation}

It is interesting to compare the small- and large-$y$ results,
to leading order in $1/d$.
As is well known, the first order in this expansion is equivalent to
any MF approximation, up to
higher-order terms.
For this purpose, we need the $1/d$ expansion of the constant $C_0$
in Eq.\ (\ref{C0}), which can be calculated as follows:
\begin{eqnarray}
C_0(d) &=& \int_{-\pi}^\pi \! \frac{d^dp}{(2\pi)^d}
 \, \frac1{\sum_{\mu=1}^d \sin^2 p_\mu} 
\ =\ 
2 \int_0^\infty \! ds \, (e^{-s} I_0(s))^d
 \label{C0exp2} \\
&=& \frac2d \, \left(1 + \frac1{2d} + {\cal O}\left(\frac1{d^2}\right) \right)
 \, , \label{C0exp3}
\end{eqnarray}
where $I_0(s) = \int_{-\pi}^\pi (d\theta/2\pi) \, \exp(s\cos\theta)$
is the modified Bessel function.
In fact, the second equality in  Eq.\ (\ref{C0exp2}) was used to obtain
the numerical result (\ref{C0}) for $C_0$.

Keeping only the leading-order term $2/d$ for $C_0$ we find that the
phase transition lines would meet at $y^2 =2/d$.

Now we are ready to map out the phase diagram of the model, as predicted
by the MF method for the weak and strong coupling regions.
This is done in Fig.\ \ref{PHASES}.
The vertical axes at $y=0$ and $y=\infty$ correspond to the O(3) model,
with its disordered (PM) and ordered (FM and AFM) phases.
These phases extend into the $y$-direction, both for $y>0$ and $y<\infty$.
Note that all the phase transition lines bend downward.
This can be understood intuitively by assuming a MF value for the
fermion condensate, which would act as an external field tending to
align the spins $\phi$ in parallel.

\section{Monte Carlo: Method and Results}
\label{MCPD}

A well established method for dynamical fermion simulations is Hybrid
Monte Carlo (HMC)~\cite{HMC}. However, the implementation of this
algorithm in a model with constrained variables is not
straightforward.  This has been satisfactorily achieved for models
with variables belonging to a Lie group~\cite{LIE}, like SU($N$) gauge
theories or like some spin-models, such as the O($N=2,4$) non-linear
$\sigma$-models. However, for other spin variables (not in a Lie group), as in the 
O(3) non-linear $\sigma$-model, this had not been satisfactorily  
solved yet, although the problem arose already in the first
simulations using the Langevin algorithm \cite{PARISI}.
Our solution is a generalization
of the strategy in~\cite{LIE}.

We shall first
discuss our solution in the quenched approximation, where comparison
with other algorithms is possible (Section \ref{QUENCHED}), 
and then deal with the full theory in Section \ref{FULL}. 
Finally our Monte Carlo results for the phase diagram of the full
theory will be presented in Section \ref{NUMERICAL}.  

\subsection{The HMC method for the quenched approximation}
\label{QUENCHED}

For the purpose of discussion it will prove convenient to briefly
describe the HMC method for unconstrained bosonic variables $\phi(x)$ with
action $S_B(\phi)$ (see ref.~\cite{HMCBOOK} for a pedagogical
presentation):

\begin{enumerate}
\item Introduce uncorrelated gaussian variables $\pi(x)$ of unit variance
(the {\it conjugate momenta} for the fields $\phi$) and
define a Hamiltonian 
\begin{equation}
H=\sum_x \frac{1}{2}\pi^2(x)+S_B(\phi)
 \, . \label{Ham}
\end{equation}
One can then use the hamiltonian equations of motion
\begin{eqnarray}
{\dot\phi}(x,\tau) &=& \pi(x,\tau) \, , \label{eom1} \label{eom2}\\
{\dot\pi}(x,\tau) &=& -\frac{\delta S_B}{\delta \phi(x,\tau)} \, , \nonumber
\label{TEQ}
\end{eqnarray}
to perform a microcanonical Molecular Dynamics evolution in
``Monte Carlo time'', $\tau$.
After a certain period of MC time (called ``trajectory''),
new random momenta $\pi(x)$ are chosen (``refreshing'' the momenta).
The crucial properties of  Eqs.\ (\ref{TEQ}) are their 
time reversibility, 
and the invariance of the Liouville measure, 
$D \phi\, D \pi$, under time evolution.
\item
In practice, the molecular dynamics equations of motion for a trajectory are
discretized into $N$ steps $\Delta\tau$.  This is done using a
leap-frog algorithm which is {\it exactly} time reversible, but does
introduce a systematic error which shows up as a non-zero $\Delta H =
{\cal O}(\Delta\tau^2)$.  The {\it detailed-balance} is not endangered
by this error, because a Metropolis acceptance step is performed.  For
fixed trajectory length, $N$ can then be tuned to optimize the overall
efficiency.

\end{enumerate}

To generalize the method to constrained variables, one needs to
appropriately define the conjugate momenta and the equations of motion
in order to preserve the constraint and, most importantly, not to
spoil the time reversibility. Each spin variable,
${\mbox{\boldmath$\phi$}}$, lives on the surface of a two-sphere, and
correspondingly one could imagine an algorithm with two independent
conjugate momenta, maybe living in the perpendicular plane
(${\mbox{\boldmath$\phi$}}\cdot{\mbox{\boldmath$\pi$}}=0$). However,
changing the constraint from the field ${\mbox{\boldmath$\phi$}}$ to
the momenta is not very appealing (and, from the practical side, one
would need to worry about {\it two} constraints in the numerical
integration).
A different approach, the use of
spherical coordinates, has the drawback of a non-planar integration
measure. Our very simple algorithm avoids constraints and non-planar
measures, by introducing {\it three} conjugate momenta per spin.

We shall start from an analogy with the
dynamics of a particle living in the sphere, a potential ($V$) acting on it.
The Hamiltonian is
\begin{equation}
H^{\mathrm{sphere}}=\frac{{\mbox{\boldmath$L$}}^2}{2} + V({\mbox{\boldmath$\phi$}}).
\label{HSPHERE}
\end{equation}
Here ${\mbox{\boldmath$L$}}$ is the angular momentum,
${\mbox{\boldmath$\phi$}}\times{\dot{\mbox{\boldmath$\phi$}}}\,$.
The equations of motion can now be obtained from the Poisson
Bracket~\cite{GOLDSTEIN} with the hamiltonian (\ref{HSPHERE}):
\begin{equation}
{\dot{{\mbox{\boldmath $\phi$}}}}\ =\ 
{{\mbox{\boldmath $L$}}}\times{{\mbox{\boldmath $\phi$}}}\ ,\ 
{\dot{{\mbox{\boldmath $L$}}}}\ =\ -{{\mbox{\boldmath $\phi$}}}\times
\frac{\delta V}{\delta {{\mbox{\boldmath $\phi$}}}}.
\label{EQSPHERE}
\end{equation} 
In this expression $\frac{\delta V}{\delta{{\mbox{\boldmath $\phi$}}}}$
stands for $\left(\frac{\delta V}{\delta\phi_1},
\frac{\delta V}{\delta\phi_2},\frac{\delta V}{\delta\phi_3}\right)$.
 
This formalism is still
inconvenient for us, because the constraint
${\mbox{\boldmath$\phi$}}\cdot{\mbox{\boldmath$L$}}=0$ complicates the
generation of random momenta according
to a Gaussian distribution.
However, the following simple facts can be straightforwardly
established from the equations (\ref{EQSPHERE}):

{\bf I.}~Both ${\mbox{\boldmath$\phi$}}^2$ and 
${\mbox{\boldmath$\phi$}}\cdot{\mbox{\boldmath$L$}}$ are conserved
through the time evolution. If the initial condition verifies the
constraints
${\mbox{\boldmath$\phi$}}^2=1\ ,\ 
{\mbox{\boldmath$\phi$}}\cdot{\mbox{\boldmath$L$}}=0$, this
will not be spoiled by the dynamics.

{\bf II.}~The dynamics is time-reversible.

{\bf III.}~Although the $L_i$ cannot be
all canonical variables~\cite{GOLDSTEIN},
the ``Liouville'' measure, 
$D {\mbox{\boldmath$\phi$}}\, D {\mbox{\boldmath$L$}}(=
d \phi_1\, d \phi_2\, d \phi_3\ d L_1\, d L_2\, d L_3)$,
is left invariant by the time-evolution.

{\bf IV.}~The Hamiltonian is a constant of the motion.\\ Now let us forget
about the constraint
${\mbox{\boldmath$\phi$}}\cdot{\mbox{\boldmath$L$}}=0$, {\em i.e.} we
introduce a new field ${\mbox{\boldmath$P$}}$ which can have a
``radial component'' (it is no longer an angular momentum), but we keep
the Eqs. of motion (\ref{EQSPHERE}). Obviously, statements {\bf
I}--{\bf IV} will still hold. Whether a symplectic structure
is hidden under this new dynamical system is unclear, but also
irrelevant (properties {\bf II} and {\bf III} are the essential ones
for HMC to be a correct algorithm~\cite{HMCBOOK}).

So, we introduce three momenta per spin, ${\mbox{\boldmath $P$}}=(P_1,P_2,P_3)$,
and write down the Hamiltonian
\begin{equation}
H \ =\ \sum_{x}\frac{{\mbox{\boldmath $P$}}^2}{2} +
       S_B({\mbox{\boldmath $\phi$}}).
\label{HQUENCHED}
\end{equation}
Equations of motion  respecting properties {\bf I}--{\bf IV} are easily generalized:
\begin{equation}
{\dot{{\mbox{\boldmath $\phi$}}}}_{(x,\tau)}  \ =\ 
   {{\mbox{\boldmath $P$}}}_{(x,\tau)}
   \times{{\mbox{\boldmath $\phi$}}}_{(x,\tau)}\  ,\ 
   {\dot{{\mbox{\boldmath $P$}}}}_{(x,\tau)}  \ =\ 
   -{{\mbox{\boldmath $\phi$}}}_{(x,\tau)}\times
   \frac{\delta S_B}{\delta{{\mbox{\boldmath $\phi$}}}_{(x,\tau)}} \, .
\label{EQQUENCHED}
\end{equation}
As expected, the evolution equations for the $S^2$ fields $\phi$
take the form of (infinitesimal) rotations, while the conjugate momenta 
can be considered as living in the Lie Algebra of SO(3).
The discretized leap-frog form of these equations is therefore naturally
formulated in terms of finite SO(3) rotations,
\begin{eqnarray}
{\mbox{\boldmath $\phi$}}_x(n\Delta\tau+\Delta\tau)&=&
\exp[\Delta\tau{\mbox{\boldmath $P$}}_x((n+\frac{1}{2})\Delta\tau)\cdot
{\mbox{\boldmath $J$}}] \,
{\mbox{\boldmath $\phi$}}_x(n\Delta\tau)
   \, , \label{eom1a}\\
{\mbox{\boldmath $P$}}_x((n+\frac{1}{2})\Delta\tau)&=&
{\mbox{\boldmath $P$}}_x((n-\frac{1}{2})\Delta\tau)\, -\,
{{\mbox{\boldmath $\phi$}}}_{(x,n\Delta\tau)}\times
\frac{\delta S_B}{\delta{{\mbox{\boldmath $\phi$}}}_{(x,n\Delta\tau)}}
     \Delta\tau
   \, , \label{eom2a}
\end{eqnarray}
where ${\mbox{\boldmath $J$}}$ are the $3\,\times\,3$ generators of SO(3),
satisfying
\begin{equation}
\left(\exp[\theta{\mbox{\boldmath $n$}}\cdot{\mbox{\boldmath $J$}}]\right)_{ij}
 \ =\ 
  \delta_{ij}\,\cos\theta+n_in_j\,(1-\cos\theta)-\epsilon_{ijk}n_k\,\sin\theta
 \label{Jchoice}
\end{equation}
for unit vectors ${\mbox{\boldmath $n$}}$.
Again, the length constraint on the $\phi$ fields is preserved by construction.

This final result is reminiscent of the elegant solution
for models with variables belonging to a Lie group and conjugate
momenta in the group algebra (or vice versa) \cite{LIE}.

\begin{table}
\caption{Values for several observables in the quenched model (\ref{ACCION})
on an $8^3$ lattice at $k=0.693 \approx k_c$, obtained with our implementation
of HMC and with Wolff's single cluster algorithm \protect\cite{WOLFF}.}
\medskip
\hrule\hrule
\begin{tabular*}{\hsize}{@{\extracolsep{\fill}}llllll}
Algorithm & \multicolumn{1}{c}{$\langle E \rangle$}
& \multicolumn{1}{c}{$\partial_k\langle E \rangle$}
&\multicolumn{1}{c}{$\chi/V$}&\multicolumn{1}{c}{$\xi$}
&\multicolumn{1}{c}{$B$}\\
\hline
HMC   & 0.3505(5) &  1.51(2)  & 0.1426(9)   & 4.47(2) & 0.800(6) \\
Wolff & 0.35061(13) & 1.501(10) & 0.1432(2) & 4.486(9) & 0.8031(18)
\end{tabular*}
\hrule\hrule
\label{TABLAQ}
\end{table}

In our case, $S_{B\, \mathrm{quenched}}=-k \sum_{n,\mu} \, {\mbox{\boldmath
$\phi$}}_n\cdot {\mbox{\boldmath $\phi$}} _{n+\hat\mu}$, so the HMC
algorithm can now be implemented in a straightforward manner.  To test the
algorithm,
we have simulated the O(3) model on an $8^3$ lattice at
$k=0.693 \approx k_c$ \cite{O3} with our
HMC algorithm and with Wolff's single-cluster embedding algorithm
\cite{WOLFF}. Let us first define the measured observables,
and then compare them.

In this work we have only measured bosonic observables, as our sole objective
was the numerical determination of the phase diagram. We have 
constructed our observables in terms of the Fourier transform
of the spin field:
\begin{equation}
\widehat{m}(\mbox{\scriptsize \boldmath $p$}) \ =\
   \frac{1}{V}\sum_{\mbox{\scriptsize \boldmath $x$}}
    \exp(-i{\mbox{\boldmath $p$}}\mathbf{\cdot} {\mbox{\boldmath $x$}})\ 
    {{\mbox{\boldmath $\phi$}}}_{\mbox{\scriptsize \boldmath $x$}} \, ,
  \label{mdef}
\end{equation}
where $V = L^3$ is the lattice volume.

We define the non-connected finite-volume susceptibilities as
\begin{equation}
{\mbox{$\chi$}}=
V \left\langle \widehat{m}^2 (0,0,0)\right\rangle , \quad
\quad {\mbox{$\chi$}}_\mathrm{s}= 
V \left\langle \widehat{m}^2 (\pi,\pi,\pi)\right\rangle.
  \label{susc}
\end{equation}

The subscript `s' on $\chi_\mathrm{s}$ stands for `staggered',
and this term is used to label quantities which are taken with a weight
$-1$ for the odd lattice sites, corresponding to momentum $(\pi,\pi,\pi)$.
Notice that
${\mbox{$\chi$}}/V$ is a pseudo order parameter, which should
be of order one in a ferromagnetically broken phase, and of
order $1/V$ in a paramagnetic or antiferromagnetic phase (and similarly for
${\mbox{$\chi$}}_\mathrm{s}/V$).

Another quantity of interest is the Binder cumulant

\begin{equation}
B=\frac{5}{2}-\frac{3}{2}
\frac{\left\langle\left( \widehat{m}^2 (0,0,0)\right)^2\right\rangle}
{\left\langle \widehat{m}^2 (0,0,0)\right\rangle^2} \, ,
  \label{Bdef}
\end{equation}
with an analogous definition for the staggered variant $B_\mathrm{s}$.

One expects $B=1$ in the FM phase, where
${\mbox{$\chi$}}/V$ is non-vanishing in the thermodynamic limit, while
it should be of order $1/V$ in the PM phase, far from the phase transition.

For the correlation length, we use a definition which 
is easy to measure and gives accurate results: 
\begin{equation}
\xi=\left(\frac{\chi/F-1}{4\sin^2(\pi/L)}\right)^{1/2},
\label{XI}
\end{equation}
where  $F$ is the squared Fourier transform at minimal 
non-zero momentum,
\begin{equation}
F \ =\ \frac{V}{3}\left(\left\langle\left | \widehat{m}
      (2\pi/L,0,0)\right |^2\right\rangle
      \ + \  \mathrm{permutations}\right) \, .
\label{secondFdef}
\end{equation}
Again, the generalization to staggered quantities is straightforward.
Another kind of observable, needed for the standard
extrapolation method~\cite{FS}, is the normalized nearest-neighbour energy
\begin{equation}
E\ =\ \frac{1}{3V}\sum_{x,\mu} \, \left\langle{\mbox{\boldmath
  $\phi$}}_x\cdot {\mbox{\boldmath $\phi$}}_{x+\hat\mu}\right\rangle
 \ =\ \frac\partial{\partial k} \ln Z \, .  \label{Enn}
\end{equation}

We also measure its fluctuation, given by 
\begin{equation}
3V \left( \left\langle E^2 \right\rangle - \left\langle E \right\rangle^2\right)
\ = \ \frac\partial{\partial k}\left\langle E \right\rangle 
 \, . \label{Ennfluct}
\end{equation}

In Table~\ref{TABLAQ} we compare the values obtained for these observables,
using our HMC algorithm and the single-cluster algorithm.
We find excellent agreement.
Of course the efficiency of our implementation of HMC is not competitive with
a cluster method in the O(3) non-linear $\sigma$-model.
But it could be useful in other
models where cluster methods are not effective in reducing the
dynamical critical exponent $z$ (for instance, when some kind of 
frustration is present \cite{NOCLUSTER}), while HMC
is expected to yield $z=1$ for any bosonic model.

\subsection{The HMC algorithm for the full theory}
\label{FULL}

The only restriction imposed on HMC is that the fermion bilinear
in the action should be given in terms of a positive definite matrix.
This will the case if we consider two identical fermion families ($N_f=2$)
as is usually done in lattice gauge theories.
After integrating them out we obtain $(\det \hat M)^2 =
\det ({\hat M}^{\dag} \hat M$),
where $\hat M$ is the fermion matrix for a single fermion family.
As we are mainly interested in the strong spin-fermion coupling region, 
it makes sense to perform the following manipulation:
\begin{equation}
\det\, \hat M \ =\ \det\, (Y+K) \ =\ y^{4V}\,\det\, (1+Y^{-1}\!K)
\label{STRONG}
\end{equation}
(cf.\ Eqs.\ (\ref{Mylarge0},\ref{Mylarge})).
The constant factor $y^{4V}$ can be dropped, and we define $M=1+Y^{-1}\!K$.

Next, one re-exponentiates the (inverse) fermion matrix
by introducing the so-called {\em pseudo-fermions\/} $z_x$,
which are complex four-component c-number fields.
The partition function is then
\begin{equation}
Z \ =\ \int\!D{\mbox{\boldmath $\phi$}} \, D\bar z \, Dz \, 
  \exp\left(-S_B-\bar z ({M}^{\dag}  M)^{-1} z\right).
\end{equation}
For further details we refer to Ref.\ \cite{HMCBOOK}.

Now the HMC Hamiltonian becomes
\begin{equation}
H=\sum_x \frac{1}{2}{\mbox{\boldmath $P$}}^2_x\ -\ k\,
\sum_{x,\mu}{\mbox{\boldmath $\phi$}}_x\cdot{\mbox{\boldmath $\phi$}}_{x+\mu}
\ +\ z^{\dag}\left(M^{\dag} M\right)^{-1} z \, ,
\label{FULLH}
\end{equation} 
and the time reversible, constraint and energy preserving equations
of motion are
\begin{eqnarray}
\label{FULLEQ}
{\dot{{\mbox{\boldmath $\phi$}}}}_{(x,\tau)}&=&{{\mbox{\boldmath $P$}}}_{(x,\tau)}\times{{\mbox{\boldmath $\phi$}}}_{(x,\tau)},\\\nonumber
{\dot{{\mbox{\boldmath $P$}}}}_{(x,\tau)}&=&
-k\sum_\mu \left({\mbox{\boldmath $\phi$}}_{(x+\mu,\tau)}+
{\mbox{\boldmath $\phi$}}_{(x-\mu,\tau)}\right)\times
{\mbox{\boldmath $\phi$}}_{(x,\tau)}\\\nonumber  & &
-z^{\dag}\left(M^{\dag} M\right)^{-1}\left[\left(
\frac{\delta M^{\dag}}{\delta{{\mbox{\boldmath $\phi$}}}_{(x,\tau)}}\times
{\mbox{\boldmath $\phi$}}_{(x,\tau)}\right)M\ +\ h.c.\ \right]
\left(M^{\dag} M\right)^{-1}z.
\end{eqnarray}

For the inversion of the fermionic matrix, we have employed 
the conjugate gradient algorithm. To formulate the stopping criterium,
let us define $h = \left(M^{\dag} M\right)^{-1}z$, $h_n$ being the 
$n^{\mathrm {th}}$ trial solution.
We continued the conjugate gradient iteration until
\begin{equation}
\frac{\left| \left(M^{\dag} M\right)h_n\ -\ z \right|^2}{|h_n|^2}\leq R.
\label{STOP}
\end{equation} 

In the simulation, we need the inverse matrix both for the leap-frog
and for the Metropolis accept-reject step. It is clear that $R$ does
not need to be the same in both cases. For the Metropolis step, lack of
accuracy in the inversion will bias the simulation. To control this,
we have checked that the Creutz parameter $\left\langle\exp(-\Delta
H)\right\rangle$ equals $1$ within errors.  In some regions of
parameter space $R$ values as small as $10^{-25}$ were needed.  The
essential requirement on the leap-frog is full reversibility in the
numerical integration of the equations of motion (up to the numerical
precision reachable with 64-bit floating point arithmetic). As first
noticed in ref.\ \cite{GUPTA}, this has no relation with $R$ if the
seed for the conjugate-gradient algorithm is chosen to depend on
the {\it actual} configuration only ($h_0=z$, for instance). However, if
$R$ is too large, the numerical integration will produce large changes
in the Hamiltonian, and the Metropolis acceptance will be poor. We
have found that $R=10^{-7}$ during the leap-frog
steps allows for a  $50\%$ acceptance.

In a first implementation of a new MC algorithm, some consistency
checks are extremely useful.  In addition, there are
three parameters to be adjusted for optimal performance, $N$,$\tau$ and
$R$. We have carried out the following tests:
\begin{enumerate} 
\item We have explicitly checked reversibility of the leap-frog algorithm.
\item We have checked that $\left\langle\exp(-\Delta H)\right\rangle = 1$
within errors.
\item The gaussian expectation values,
$\left\langle z^{\dag}\left(M^{\dag} M\right)^{-1}z\right\rangle=4$ and 
$\left\langle{\mbox{\boldmath $P$}}^2\right\rangle=3$ have been checked.
\item We have checked that $\Delta H\propto (\Delta\tau)^2$ in the leap-frog
integration, for constant trajectory length $N\Delta\tau$.
\end{enumerate} 

In addition, 
we compared 
simulation results for the full theory at $(k,y)$, with the output of
a quenched simulation at the corresponding effective coupling value
obtained in a large-$y$ expansion,
\begin{equation}
k^{{\mathrm {eff}}}=k+\frac{2}{y^2}+O\left(\frac{1}{y^4}\right)
\label{keffagain}
\end{equation}
(cf.\ Eq.\ (\ref{kRenorm})).
In table (\ref{TABLATEST}), we give the mean value of several operators
as obtained on a $4^3$ lattice at $k=0.693$, $y=10.0$ and in the
quenched theory. The agreement is excellent. 
Notice that even if the shift in 
the effective coupling is only $3\%$, the effects of the dynamical fermions 
can be clearly measured as  the observables change quite 
significantly  at the critical point $k_{\mathrm c}=0.693$.

\begin{table}[b]
\caption{Comparison of observables in the full
theory (\ref{ACCION}) at $(k = 0.693,y = 10.0)$
and in the quenched model both at the corresponding value of
$k^{{\mathrm {eff}}}$ and at $k_c = 0.693$. 
We have $140,000$ unquenched trajectories ($N$=10, $\Delta\tau$=0.3)
on a $4^3$ lattice. The Metropolis acceptance rate was  65-70\%,
with an autocorrelation time of 3-4 trajectories.}
\medskip
\hrule\hrule
\begin{tabular*}{\hsize}{@{\extracolsep{\fill}}lllll}
Couplings & \multicolumn{1}{c}{$\langle E \rangle$}
& \multicolumn{1}{c}{$\partial_k\langle E \rangle$}
&\multicolumn{1}{c}{$\chi/V$}&\multicolumn{1}{c}{$\xi$}\\\hline
$k$=0.693 , $y$=10.0    & 0.4164(6) &  1.134(6)  & 0.3111(7)   & 2.378(4)\\
$k$=0.713 , $y$=0       & 0.41584(14) &  1.130(4)  & 0.3108(2)   & 2.3779(18)\\
$k$=0.693 , $y$=0       & 0.3928(3) &  1.174(4)  & 0.2836(4)   & 2.214(2)
\end{tabular*}
\label{TABLATEST}
\hrule\hrule
\end{table}

\subsection{Phase Diagram}
\label{NUMERICAL}

\begin{figure}[htb]
\begin{center}
\leavevmode
\centering\epsfig{file=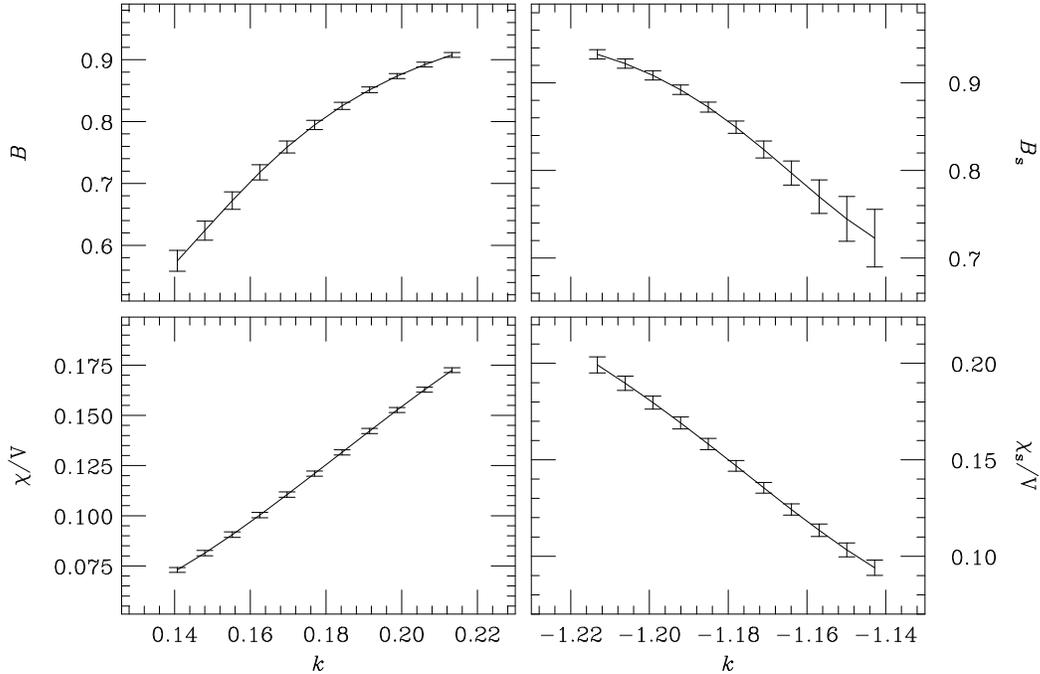,width=0.6\linewidth,angle=90}
\end{center}
\caption{
Binder cumulant (\protect\ref{Bdef}) and non-connected susceptibility
(\protect\ref{susc}) as a function of $k$, around the two critical
points at $y=2.0$. For each critical point, only one simulation
has been carried out. The other points are obtained with the 
standard reweighting method.}
\label{GRANY}
\end{figure}

\begin{figure}[htb]
\begin{center}
\leavevmode
\centering\epsfig{file=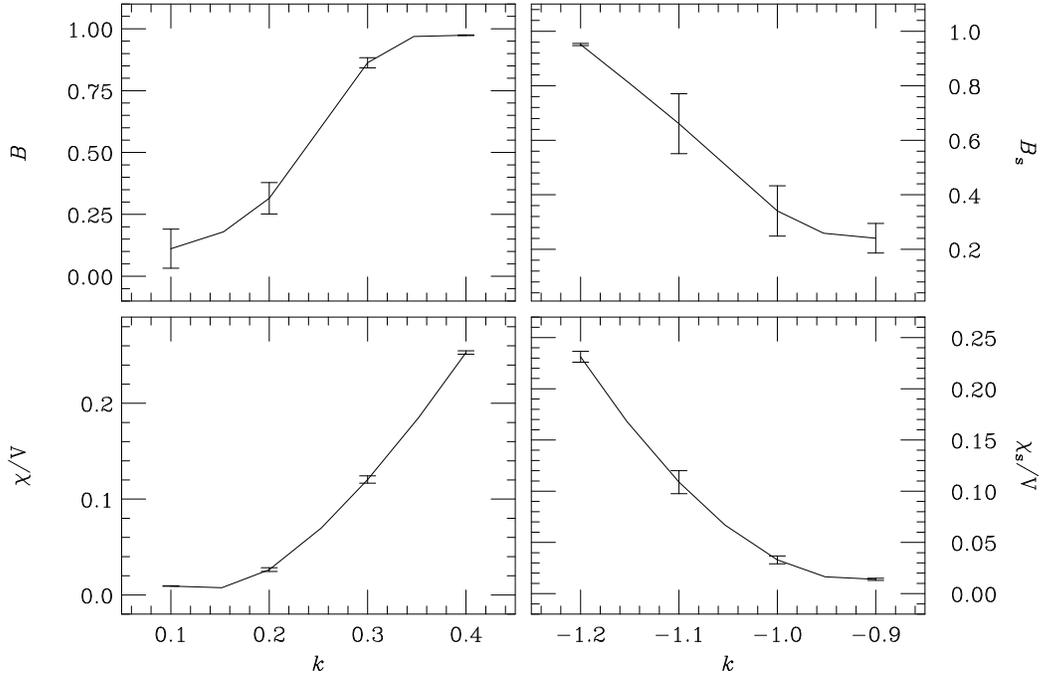,width=0.6\linewidth,angle=90}
\end{center}
\caption{
Binder cumulant (\protect\ref{Bdef}) and non-connected susceptibility
(\protect\ref{susc}) as a function of $k$, around the two critical
points at $y=0.5$. The data points are from different simulations.
}
\label{SMALLY}
\end{figure}

\begin{figure}[htb]
\begin{center}
\leavevmode
\centering\epsfig{file=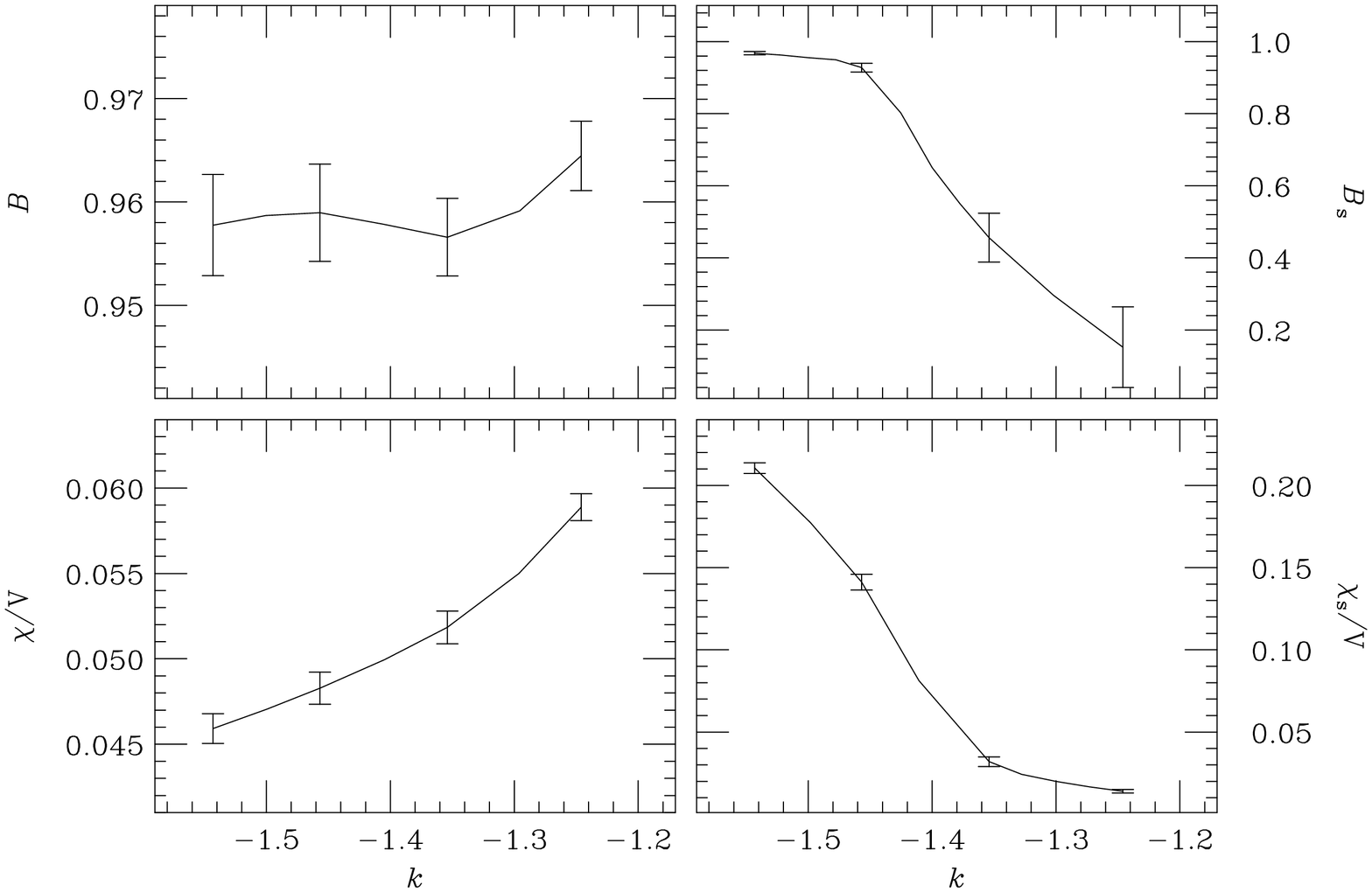,width=0.6\linewidth,angle=90}
\end{center}
\caption{
Binder cumulants and susceptibilities when crossing the FM(S)-EM
transition line at $y=1.15$.
}
\label{EMV}
\end{figure}

\begin{figure}[htb]
\begin{center}
\leavevmode
\centering\epsfig{file=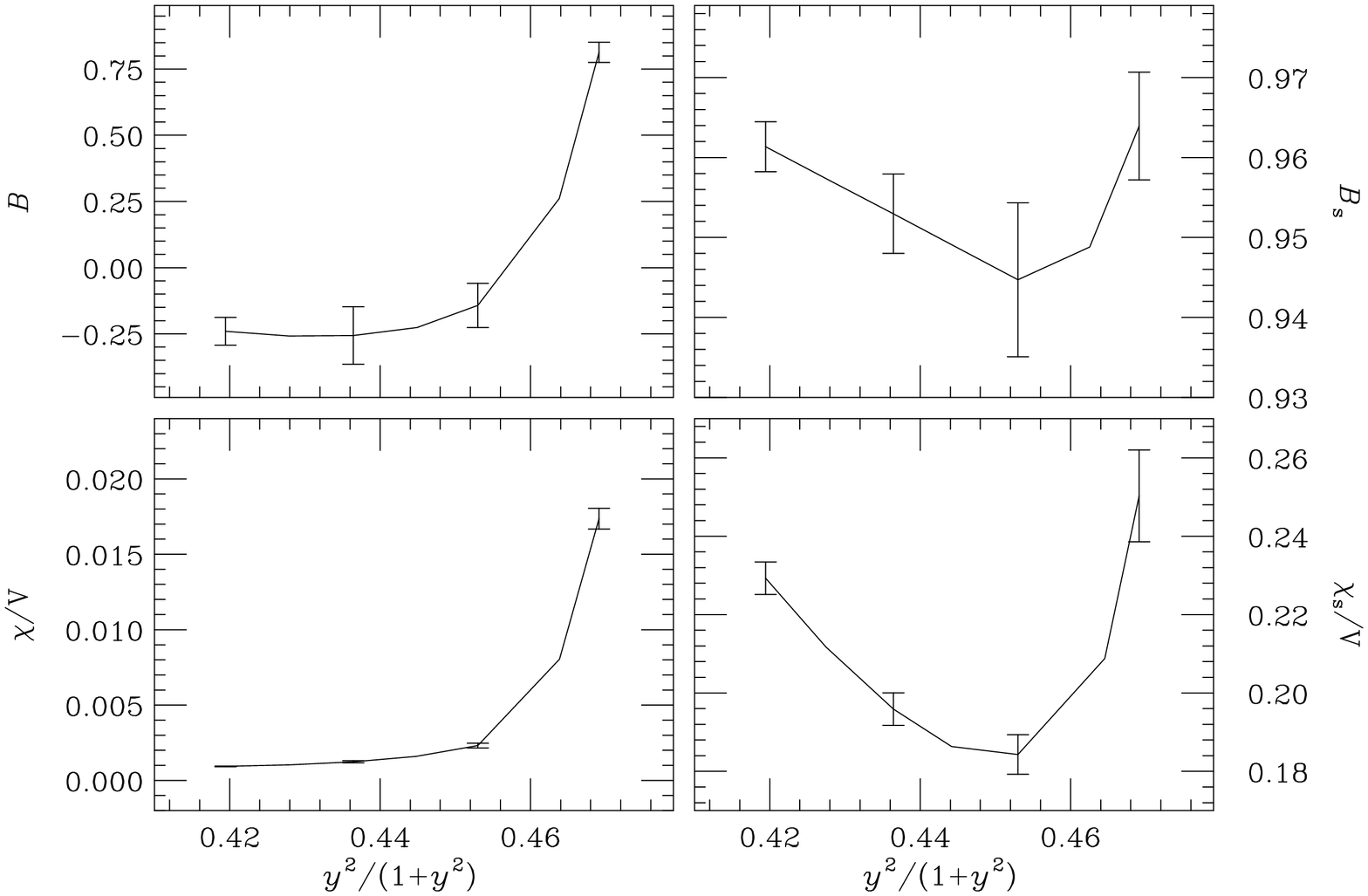,width=0.6\linewidth,angle=90}
\end{center}
\caption{
Binder cumulants and susceptibilities when crossing the AFM(W)-EM
transition line at $k=-1.6$.
}
\label{EMH}
\end{figure}

The phase diagram in fig.\ \ref{PHASES}  was obtained on an $8^3$ lattice.
As there is no true phase transition on a finite lattice, a criterium
is needed to locate the phase boundaries. We looked for the point
where the relevant Binder cumulant equals the value $B=0.8$ it has at
($k=\pm 0.693 \approx k_\mathrm{c}$, $y=0$).
Since $B=1$ deep in the broken phase and
$B\propto 1/L^3$ in the symmetric one, this provides a clean
quantitative criterium which yields a point definitely inside the
critical region. The width of the critical region decreases as
$L^{-1/\nu}$, therefore the systematic error in the critical coupling
will be at most of order $10^{-1}$. However, since the Binder parameter is a
universal quantity, which should stay constant along much of the critical
lines, the error rather goes as $L^{-(\omega+1/\nu)}$ ({\em i.e.} 
${\cal O} (10^{-2})$). Thus, this systematic error is under control in the
full theory as well.  
We used the standard reweighting method~\cite{FS} to
determine the precise location of the points where $B=0.8$.

The total simulation time was 16 days of the 32 Pentium Pro processor parallel
computer RTNN based in Zaragoza. To allow for a correct
thermalization, we discarded $100$ integrated autocorrelation times of the
relevant susceptibility. This may look utterly conservative, and the
MC history indeed seems to stabilize long before that. However, not
much is known about the {\it exponential} autocorrelation time
of fermionic algorithms and one should be cautious.

As Eq.\ (\ref{STRONG}) shows, both at $y=\infty$ and at $y=0$ we
recover the non-linear $\sigma$-model with its well known
paramagnetic, ferromagnetic and antiferromagnetic phases. At finite
$y$, we expect these phases to extend into the ($k$,$y$) plane.
In fact one can quite precisely anticipate the critical coupling from the
strong coupling formula (\ref{keffagain})
and the quenched critical points $k_{\mathrm c}^{(y=\infty)}=\pm 0.693$.
Using the reweighting method, the phase
transition lines can be determined down to $y\approx 2.0$.
In fig.\ \ref{GRANY} the variation of the
Binder cumulant and the susceptibility around the two critical
couplings is shown for $y=2.0$.

In the small-$y$ region, the effective action up to ${\cal O}(y^2)$ does
not only renormalize $k$, but also introduces additional couplings,
due to the non-locality of the matrix $K^{-1}$ occurring in the weak-coupling
expansion.
Therefore, we do not have an estimate for $k^{{\mathrm {eff}}}$ as reliable
as in the large-$y$ region (\ref{keffagain}), but we can nevertheless
obtain an estimate for $k_{\mathrm c}(y)$ from the MF approximation.
We have simulated at several values of
the coupling $k$, for fixed $y$, until the corresponding 
Binder parameter crossed its critical value.
A more accurate result for the critical point was later on obtained with the
reweighting method. In fig.\ \ref{SMALLY},
we have plotted the relevant Binder parameter and susceptibility for
$k$ values near the two critical couplings with $y=0.5$.

In fig.\ \ref{EMV} we show
the variation of both order parameters and Binder cumulants
when crossing the FM(S)-EM transition line at $y=1.15$. We
find a strong change in the staggered quantities, while
the non-staggered ones show a smoother evolution. However,
the non-staggered order parameter is much smaller than its
staggered counterpart. This may indicate that, although
the non-staggered sector is non-critical ($B\sim 1$),
it will eventually undergo a phase transition at lower $k$.
A similar behaviour is found when traversing the AFM(W)-EM line at
$k=-1.6$ (see fig.\ \ref{EMH}), but now the non-staggered quantities
show a more pronounced signal. The detailed study of these transition
lines (order of the phase transitions, critical exponents, etc.) 
requires a finite-size scaling analysis, which is left for future work. 
This study will be much easier if the transition line is crossed
varying $k$, as we lack an analogue of the reweighting
method for $y$.

\section{Quasiparticle excitations at the MF level}

\label{sectMFexc}

In this section we explore the relevant excitations involving
fermions, with emphasis on the strong-coupling region of our model.

The small-$y$ regime has been studied in relation with
the mechanism by which leptons and quarks acquire
their mass through symmetry breaking in the Electroweak sector of the
Standard Model. Due to the weak coupling there are no surprises.
This situation will change dramatically when we consider the strong-coupling
region, though.

\subsection{Fermionic excitations in the FM(S) and PMS phases}
\label{sectPMSexc}

At very large $y$, it is natural to attempt a large-$y$ expansion.
This can be achieved after carrying out the following change of variables:
\begin{eqnarray}
\bar\psi'&=&\bar\psi \, , \label{chofvar1} \\
\psi'&=&\left({\mbox{\boldmath $\phi$}}
       \cdot{\mbox{\boldmath $\tau$}}\right)\psi 
    \label{chofvar2} \, .
\end{eqnarray}
Because of the constraint ${\mbox{\boldmath $\phi$}}^2=1$ and the
identity
$\left({\mbox{\boldmath $\phi$}}\cdot{\mbox{\boldmath $\tau$}}\right)^2=
{\mbox{\boldmath $\phi$}}^2{\mbox{\boldmath $1$}}$, this transformation
has unit Jacobian and its inverse satisfies
\begin{equation}
\psi=\left({\mbox{\boldmath $\phi$}}\cdot{\mbox{\boldmath $\tau$}}\right)\psi'
  \, . \label{chofvarinv}
\end{equation}
In terms of the new variables (dropping the primes) the action takes the form
\begin{equation}
S= -k\, \sum_{x,\mu}\, {\mbox{\boldmath $\phi$}}_x
   \cdot{\mbox{\boldmath $\phi$}}_{x+\mu}+\sum_{x,y}\,
   \bar\psi_x\left(K_{xy}\left({\mbox{\boldmath $\phi$}}_y
   \cdot{\mbox{\boldmath $\tau$}}\right) +y\delta_{xy}\right)\psi_y,
\label{Sagain}
\end{equation}
where the fermion kinetic term is the usual lattice kinetic Dirac operator,
defined in Eq.\ (\ref{MATRIZK}).
After a further rescaling of the $\psi$ fields, the coupling $y$ can be
moved to the kinetic term, where it appears as $1/y$.

Note that this change of variables (\ref{chofvar1},\ref{chofvar2}) was
implicitly present in the MF calculations of the phase diagram in the
strong-coupling region as well (cf.\ Eqs.\
(\ref{Mylarge0},\ref{Mylarge})).  This transformation is the reason
that explains that the model is (partly) analytically
tractable. The interest of finding reliable analytical
approaches to strongly coupled fermion systems need not to be
stressed.

The fermion propagator $\langle \psi_x \bar\psi_y \rangle$ is given by
the expectation value of the inverse fermion matrix,
which in a large-$y$ expansion becomes
\begin{equation}
\langle \psi_x \bar\psi_y \rangle \ = \ \left\langle M^{-1}_{xy}\right\rangle
 \ = \ \left\langle \frac1y \left(1 - \frac1y K ({\mbox{\boldmath
 $\phi\cdot\tau$}})+ \frac{1}{y^2} K ({\mbox{\boldmath
 $\phi\cdot\tau$}}) 
K ({\mbox{\boldmath $\phi\cdot\tau$}}) - \ldots
  \right)_{xy}\right\rangle
   \, .  \label{fermprop}
\end{equation}
This can be viewed as a sum over paths of increasing length connecting
$x$ and $y$ (recall that $K$ is a nearest-neighbour matrix).

In the FM(S) phase, there is a non-zero magnetization
$v = |\langle {\mbox{\boldmath $\phi$}} \rangle|$.
Expectation values of products of ${\mbox{\boldmath $\phi$}}$ fields on different sites
are replaced by the appropriate powers of $v$.
Corrections to this approximation as well as contributions from paths
visiting a given site more than once are of higher order in $1/d$ and
are ignored at the MF level.
The series (\ref{fermprop}) can thus be resummed and one finds a
propagator
\begin{equation}
\langle \psi \bar\psi \rangle_{FM(S)} \ = \ 
\frac{1/v}{K + y/v}
\end{equation}
which is that of a fermion with a mass $y/v$.
Note that, since $v<1$, this is a huge mass if $y$ is large.
The propagator for the original fermion, before the change of variables
(\ref{chofvar1},\ref{chofvar2}), corresponds to the same physical particle;
the only difference is in the wave function renormalization.

In the PM(S) phase, $v=0$, so at the MF level the fermion would be
infinitely massive, or in other words, non-propagating.
Beyond this naive MF level, however, a large but finite mass will be found.
This is due to the next-to-leading contributions to the series
(\ref{fermprop}).
The dominant terms are now those involving the expectation value for
the nearest-neighbour energy $z^2 \equiv \langle {\mbox{\boldmath
$\phi$}}_x \cdot {\mbox{\boldmath $\phi$}}_{x+\hat\mu}
\rangle$, which is of order $1/2d$ and therefore absent at the MF level.
The resummation of contributions in (\ref{fermprop}) now leads to a
fermion propagator with a mass $y/z$, which is even larger than 
the mass of the fermion in the FM(S) phase.

The conclusion of this analysis, which is similar to that in (chiral) Yukawa
models in the Electroweak theory \cite{SMITz2}, is that the elementary
fermion excitations in the large-$y$ region are very heavy (hence essentially
non-propagating), and therefore
play no role in the spectrum of light excitations.
This holds even stronger in the PMS phase than in the FM(S) phase.

\subsection{Fermionic excitations in the AFM(S) phase}
\label{AFMSsection}

Here our point of departure is again the form of the action (\ref{Sagain}),
which is tailored for studying the large-$y$ behaviour.
In the AFM(S) phase, we have a staggered expectation value for the
${\mbox{\boldmath $\phi$}}$ field at the MF level, which can be taken in the 3-direction,
\begin{equation}
{\mbox{\boldmath $\phi$}}_x=\epsilon_x v\left(\begin{array}{c}
0\\0\\1\end{array}\right)
\label{MSTAG}
\end{equation}
(with $\epsilon_x=(-1)^{x_1+x_2+x_3}$).
Hence
\begin{equation}
\left({\mbox{\boldmath $\phi$}}_x\cdot{\mbox{\boldmath $\tau$}}\right)\psi_x=
\left(\begin{array}{r} v\,\epsilon_x\,\psi_x^{(1)}\\
-v\,\epsilon_x\,\psi_x^{(2)}\end{array}\right),
\end{equation}
where $\psi_x^{(i)}$, $i=1,2$ labels the two {\it flavours} in $\psi_x$. So
after the change of variables (\ref{chofvar1},\ref{chofvar2})
the kinetic operator in (\ref{Sagain}) is still diagonal in flavour.
The only effect of the new variables is to change the lattice Dirac
operator from (\ref{MATRIZK}) to
$$ v\,\epsilon_y\, \tau_3 K_{xy}.$$
Due to this diagonal structure in flavour space, we can concentrate on
one flavour, say $\psi^{(1)}$; the other flavour is obtained by taking $-v$
instead of $v$.
In Fourier space, the kinetic term for $\psi^{(1)}$ is given by
\begin{equation}
   -i\,v\,\ 
      {\slash\hspace{-0.8em}{\sin}}\, p\ \delta_{p,q\pm \pi},
\end{equation}
where
\begin{eqnarray}
{\slash\hspace{-0.8em}{\sin}}\, p&=&\sum_{\mu}\,\sigma_{\mu}\ \sin\,p_{\mu}
 \, , \label{sinp} \\
 \delta_{p,q\pm \pi}&=&\prod_\mu \delta_{p_\mu,q_\mu+\pi \ \mathrm{mod}\ 2\pi}
   \, . \label{deltapq}
\end{eqnarray}
So we obtain for the inverse of the MF propagator
in the AFM(S) phase, 
\begin{equation}
M_{p,q}=-i\, v\ \,{\slash\hspace{-0.8em}{\sin}}\, p\ \delta_{p,q\pm \pi}+
y\,\delta_{p,q},
\label{PROAFMRAW}
\end{equation}
or, in matrix notation for the subspace of the modes coupled in 
Eq.\ (\ref{PROAFMRAW}), $p$ and $p\pm(\pi,\pi,\pi)$,
\begin{equation}
M_{p,p\pm(\pi,\pi,\pi)}=\left(\begin{array}{cc} 
y & -i\, v\  {\slash\hspace{-0.8em}{\sin}} p\\
i\, v\  {\slash\hspace{-0.8em}{\sin}} p & y
\end{array}\right).
\label{PROAFMMAT}
\end{equation}

To find the quasiparticle excitations in the AFM(S) phase we diagonalize
the fermionic part of the action (\ref{PROAFMMAT}).
One obtains
\begin{equation}
S=\int_p \bar\psi(p) \,
(y-v\ \; {\slash\hspace{-0.8em}{\sin}}\, p) \, \psi(p)\ ,
\label{SDIAG}
\end{equation}
where
$$\psi(p)=\frac{1}{\sqrt{2}}\left[\psi^{(1)}(p)\,+
\, i\,\psi^{(1)}(p+\pi)\right],$$
or, in {\em position} space,
$$\psi_x=\frac{1}{\sqrt{2}}\left[\psi^{(1)}_x\,+
\, i\,\epsilon_x\psi^{(1)}_x\right].$$

The momentum space propagator corresponding to (\ref{SDIAG}) is thus
\begin{equation}
S(p)=\frac{1}{y-v\ \; {\slash\hspace{-0.8em}{\sin}}\, p}=
\frac{y+v\ \; {\slash\hspace{-0.8em}{\sin}}\, p}
{y^2-v^2\,\sum_{\lambda} \sin^2\, p_{\lambda}}.
\label{PROAFM}
\end{equation}
Since we are working in imaginary time, one would expect quasiparticle poles
in $S(p)$ to appear at negative values of $p^2$.
The unusual relative minus sign in the denominator (\ref{PROAFM}) therefore
does not seem to allow for a quasiparticle interpretation, at first sight.

However, (\ref{PROAFM}) suggests the possibility of
light excitations with a relativistic dispersion relation around 
momenta $\left(\pm\frac{\pi}{2},\pm\frac{\pi}{2},\pm\frac{\pi}{2}\right)$.
To see this, consider the denominator in Eq.\ (\ref{PROAFM})
for small $k_\mu = p_\mu \pm \pi/2$:
\begin{equation}
y^2-v^2\,\sum_{\lambda} \sin^2\, p_{\lambda}\ =\ (y^2-v^2\, d)\ +\ 
v^2\,\sum_{\lambda}\,k_{\lambda}^2\ +\ {\cal O}(k^4),
\label{TAYLOR}
\end{equation}
where $d=3$ is the space-time dimension.
As long as we are at large enough $y$, such that $y^2\,>\, d\,v^2$
(recall $v^2<1$), this dispersion relation corresponds to a
relativistic excitation with $m^2=(y^2-d v^2)/v^2$, in this naive
MF calculation.
Several comments are in order:

\begin{enumerate}
\item For $v$=0, we recover the MF result for the PMS phase:
the kinetic term in (\ref{TAYLOR}) is suppressed.
\item At the MF level, only for $(y^2-dv^2)$ small enough compared to
$v^2$ these fermionic excitations,
$\left({\mbox{\boldmath $\tau$}}\cdot{\mbox{\boldmath $\phi$}}\right)\psi$,
can propagate easily. Since $v^2<1$, this can only happen for $y^2$
not too large.
\item These would-be excitations are characteristic of the AFM(S) phase.
Let us recall that in the PMS phase no {\it light} fermionic excitations have 
been identified at the MF level.

\end{enumerate}

\subsection{Light bound states in the PMS phase}
\label{excPMSbos}

\begin{figure}[htb]
\begin{center}
\leavevmode
\centering\epsfig{file=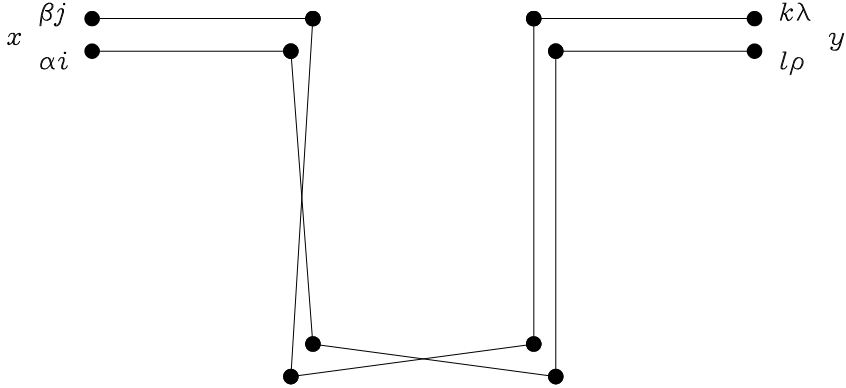,width=0.35\linewidth,angle=90}
\end{center}
\caption{
A typical double-chain diagram, connecting sites $x$ and $y$.
The chains are parallel in position space.
}
\label{DOUBLECHAIN}
\end{figure}

We have seen above that the fermionic excitations in the PMS phase are
very heavy.
We will now show that there are bound states of elementary fermions,
however, which are light.
This is done by means of a MF calculation of the double-chain type
\cite{STEPHANOV}.

Consider the propagator for a fermion pair $\psi_x\psi_x$,
\begin{equation}
 \langle \psi^\alpha_{x,i} \psi^\beta_{x,j}
     \bar\psi^\lambda_{y,k} \bar\psi^\rho_{y,l} \rangle
\ = \
\langle M^{-1}_{x,\beta,j;\,y,\lambda,k} \, M^{-1}_{x,\alpha,i;\,y,\rho,l} \rangle -
\langle M^{-1}_{x,\alpha,i;\,y,\lambda,k} \,
     M^{-1}_{x,\beta,j;\,y,\rho,l} \rangle\ .
\label{pairprop}
\end{equation}
Here $M^{-1}$ is the single-fermion propagator,
$\alpha, \beta, \lambda, \rho$ are Dirac indices, and $i, j, k, l$
are flavour indices. Thus, this propagator is really a $16\times 16$ matrix.
For the moment we keep all these indices as they are; later on we will discuss
how pairs of them decompose into quantum numbers for the composite state.

Let us concentrate on the first $\langle M^{-1} M^{-1} \rangle$ term in
Eq.\ (\ref{pairprop}).
Using the $1/y$ expansion of $M^{-1}$ as before, we find the series
\begin{equation}
\langle M^{-1}_{x,\beta,j;\,y,\lambda,k} \, M^{-1}_{x,\alpha,i;\,y,\rho,l} \rangle
 \ = \ 
\sum_{N,N'=0}^\infty \left\langle \left[ \frac{\phi}{y} \left( K\frac{\phi}{y}
  \right)^N \right]_{x,\beta,j;\,y,\lambda,k}
 \left[ \frac{\phi}{y} \left( K\frac{\phi}{y}
  \right)^{N'} \right]_{x,\alpha,i;\,y,\rho,l} \right\rangle
 \, , \label{pairseries}
\end{equation}
where we have written $\phi$ as a shorthand for 
$({\mbox{\boldmath $\phi$}}\cdot{\mbox{\boldmath $\tau$}})$. It is clear
that 
only terms with $N+N'$ even survive in a paramagnetic phase, due
to the ${\mbox{\boldmath $\phi$}}\rightarrow -{\mbox{\boldmath $\phi$}}$ symmetry, thus a factor $(-1)^{N+N'}$
has been dropped.
Since the matrix $K$ connects nearest-neighbour sites only, each term
in this series can be seen to represent a product of two paths (chains) of
lengths $N$ and $N'$ respectively, connecting site $x$ with site $y$
(so, if the ``distance'' between $x$  and $y$ is even(odd), both $N$ and
$N'$ will be even(odd)).

We will attempt to sum the complete series, to leading order in $1/d$,
where $d=1+2=3$ is the Euclidean space-time dimension. For this, we
need the spin-spin propagator, which in this approximation
is extremely short ranged
\begin{equation}
\langle \phi_x^a \phi_{x}^b \rangle = \frac{1}{3} \delta^{ab}
 \, .
\label{SHORT}
\end{equation}
Expectation values of the type $\langle {\mbox{\boldmath $\phi$}}_x \cdot {\mbox{\boldmath $\phi$}}_{x+\hat\mu} \rangle$
are of order $1/d$, and others are suppressed even stronger.
Thus, assuming (\ref{SHORT}),
we observe that any term in the series which contains ${\mbox{\boldmath $\phi$}}_x$ for
a given site $x$ only once or an odd number of times will vanish due to
$\langle {\mbox{\boldmath $\phi$}} \rangle = 0$.
When the site is visited twice, it follows from ${\mbox{\boldmath $\phi$}}^2 = 1$ that
the contribution from the ${\mbox{\boldmath $\phi$}}$ fields
is proportional to $\frac{1}{3} \delta^{ab}$.
Thus each site along the chains connecting $x$ and $y$ must be visited
an even number of times.
One class of diagrams fulfilling this requirement consists of the
so-called `double-chain' diagrams, where the propagation of both fermions
between $x$ and $y$
follows the same path in position space (see figure \ref{DOUBLECHAIN}).
As was convincingly argued in Ref.\ \cite{STEPHANOV}, this class
saturates the dominant diagrams in the $1/d$ expansion.
Indeed, one can easily check by concrete examples, how deviations from
double-chain behaviour induce additional powers of $1/d$.
We shall also assume that these double chains are self-avoiding (this
is allowed at first order in $1/d$).

\begin{figure}[htb]
\begin{center}
\leavevmode
\centering\epsfig{file=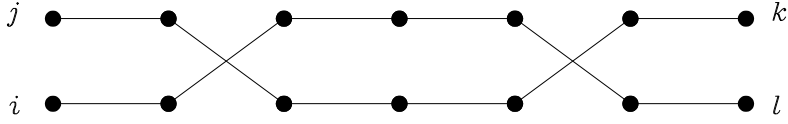,width=0.11\linewidth,angle=90}
\end{center}
\caption{
A matrix-product term contributing to the flavour structure.
}
\label{FLAVOUR}
\end{figure}

Our task is thus to sum up all double chain diagrams connecting $x$ and $y$.
Let us first consider the flavour structure.  
Using  (\ref{SHORT}) one finds that
\begin{equation}
\langle ({\mbox{\boldmath $\phi$}}_x\cdot{\mbox{\boldmath $\tau$}})_{jk} ({\mbox{\boldmath $\phi$}}_x\cdot{\mbox{\boldmath $\tau$}})_{il} \rangle
\ = \ \frac13 \sum_a \tau^a_{jk} \tau^a_{il}
\ = \ \frac13 (- \delta_{jk} \delta_{il} + 2 \delta_{jl} \delta_{ik}) 
 \, . \label{phitauphitau}
\end{equation}
>From this and from the ultra-local correlations
we are considering (cf.\  Eq.\  (\ref{SHORT})), it follows that
the product of $2(N+1)$ factors of $({\mbox{\boldmath $\phi$}}\cdot{\mbox{\boldmath $\tau$}})$
along a double chain of length $N$ visiting the points $x=x_0, x_1,\ldots,
y=x_N$ (cf.\ Eq.\ (\ref{pairseries})) is
\begin{equation}
\left\langle
   \left[
     \prod_{n=0}^N ({\mbox{\boldmath $\phi$}}_{x_n}\cdot{\mbox{\boldmath $\tau$}}) 
   \right]_{x,j;\,y,k}
  \,
   \left[
     \prod_{n'=0}^N ({\mbox{\boldmath $\phi$}}_{x_{n'}}\cdot{\mbox{\boldmath $\tau$}}) 
   \right]_{x,i;\,y,l}
\right\rangle
\ = \
P\,\delta_{jk} \delta_{il}\  + \ Q\,\delta_{jl} \delta_{ik} . 
\end{equation}
To calculate $P$ and $Q$, it is convenient to represent the general term 
contributing to the above matrix product as in figure \ref{FLAVOUR}. A
graph contributing to $\delta_{jk} \delta_{il}$ will have an even number
of crossings, while diagrams contributing to $\delta_{jl} \delta_{ik}$
jump an odd number of times. Each crossing contributes a factor $\frac{2}{3}$,
while non-crossings yield factors $-\frac{1}{3}$ (cf.\  Eq.\  (\ref{phitauphitau})).
Now, $P$ and $Q$ can be easily obtained using binomial summation:
\begin{eqnarray}
&&\left\langle
   \left[
     \prod_{n=0}^N ({\mbox{\boldmath $\phi$}}_{x_n}\cdot{\mbox{\boldmath $\tau$}}) 
   \right]_{x,j;\,y,k}
  \,
   \left[
     \prod_{n'=0}^N ({\mbox{\boldmath $\phi$}}_{x_{n'}}\cdot{\mbox{\boldmath $\tau$}}) 
   \right]_{x,i;\,y,l}
\right\rangle
  \nonumber \\
&&\ \ \ \ \ \ \ \ \ \ \ \ \ \ \  \ = \
\left( \frac13 \right)^N \,
      \frac12 ( \delta_{jk} \delta_{il} + \delta_{jl} \delta_{ik} )
 \ + \
(-1)^N \, \frac12 ( \delta_{jk} \delta_{il} - \delta_{jl} \delta_{ik} )
 \, , \label{phitauN}
\end{eqnarray}
where we have separated in a term symmetric under
$(ji)\leftrightarrow (ij)$ and an antisymmetric one (this will be needed
for separating the contribution to different quantum numbers).
It is remarkable that the flavour contribution only depends on the
double-chain length, but not on its shape. This allows for a total
factorization between flavour and Dirac indices.

Next, consider the Dirac structure.
One gets products of matrices
\begin{equation}
K^{\mu_n}_{x_n x_{n+1}} K^{\mu_n}_{x_n x_{n+1}}
\sigma^{\mu_n}_{\beta_n\lambda_n}   \sigma^{\mu_n}_{\beta_{n+1} \lambda_{n+1}},
 \label{KKss}
\end{equation}
along the double chain,
where 
\begin{equation}
K^\mu_{xy} \ = \ 
  \frac12 (\delta_{y,x+\hat\mu} - \delta_{y,x-\hat\mu})
 \, . \label{Kagain}
\end{equation}
One readily finds that
\begin{equation}
  A_{xy}^\mu\equiv 4\, K^\mu_{xy}K^\mu_{xy} \ = \ (\delta_{y,x+\hat\mu} \ + \ \delta_{y,x-\hat\mu})
 \, . \label{Amudef}
\end{equation}
Thus, we need to calculate
\begin{equation}
\sum_{\{\mu_n\}} \left[\prod_n\ {\frac{1}{4}} A^{\mu_n}\, 
\sigma^{\mu_n}\otimes\sigma^{\mu_n}\right]_{x,\beta , \alpha ; y, \lambda ,\rho
}\ ,
\label{AMUNN}
\end{equation}
where the sum is extended to all the lattice paths (denoted by $\{\mu_n\}$)  of-length $N$ connecting
$x$ and $y$. Now, we can extend the sum to {\it all} length-$N$
lattice paths starting at $x$, 
because paths not connecting $x$ to $y$ will contribute a zero $x\, y$ entry. 
This can be also understood   by realizing
that once the chain has arrived at $x_i$, there are $2d$ possible
directions to continue the chain.
These are added up by summing Eq.\ (\ref{KKss}) over $\mu$.
At the next site, we do the same for the next step along the chain.
The contributions of all double chains are therefore added up when
we take the product of these $\mu$-sums along the chain.
Corrections due to backtracking ($2d \rightarrow 2d-1$) are down by $1/d$.

So we need to calculate powers of the matrix
\begin{equation}
\frac{1}{4}
\sum_\mu A^\mu\, 
\sigma^\mu\otimes\sigma^\mu
  \, . \label{sumKKss}
\end{equation}
One way to do that is to write it out explicitly as a $4\,\times\,4$ matrix in
the space spanned by the vectors $(\beta,\lambda) = $(1,1), (2,2), (1,2)
and (2,1), in that order.
One finds that it equals
\begin{equation}
    \frac{1}{4}
    \left(
  \begin{array}{cccc}
       A^3 & A^1 - A^2 & 0 & 0 \\
      A^1 - A^2 &  A^3 & 0 & 0 \\
      0 & 0 & -A^3 & A^1 + A^2 \\
      0 & 0 & A^1 + A^2 & -A^3 
  \end{array}
    \right)\ .
\label{Amatrix}
\end{equation}
It can be diagonalized in this $4\,\times\,4$ space.
The eigenvalues, up to the factor $1/4$, are found to be
\begin{eqnarray}
&&\lambda^{\mu} \ = \ A \ - \ 2 A^{\mu} \ \ \ \ \ \ \ \ \ \ \ \ (\mu=1,2,3)
  \label{lambda123}\, , \\
&&\lambda^4 \ = \ - A\, ,
  \label{lambda4}
\end{eqnarray}
where
\begin{equation}
A \ = \ \sum_{\mu=1}^3 A^\mu  \ = \  \Box + 2d
 \, , \label{Adef}
\end{equation}
and $\Box$ is the lattice discretization of the d'Alembertian
$\sum_\mu \partial_\mu \partial_\mu$.
The $N^\mathrm{th}$ power (see (\ref{AMUNN})) of the matrix (\ref{sumKKss}) is now easy to calculate.

In order to collect the factors and sum up the contributions, let us
go back to Eq.\ (\ref{pairprop}).
We see that we need to antisymmetrize each term in $\langle M^{-1} M^{-1}
\rangle$ with respect to the simultaneous interchange of
$\alpha, i$ with $\beta, j$.
This gives a sum of two terms, one symmetric in $\alpha \leftrightarrow \beta$
and antisymmetric in $i \leftrightarrow j$ (corresponding to a composite
state which is a Dirac vector and a flavour singlet), and one vice versa
(singlet in Dirac space, vector in flavour space).
Note that Eq.\ (\ref{phitauN}) has already been written as a sum of
symmetric and antisymmetric terms.
The symmetric and antisymmetric parts of the Dirac structure correspond
to Eqs.\ (\ref{lambda123}) and (\ref{lambda4}), respectively.

Collecting the various factors, we can carry out the geometric sum over $N$ in
Eq.\ (\ref{pairprop}) and we find
the following propagators for the composite states:
\begin{itemize}
\item
a Dirac vector -- flavour singlet with propagator
\begin{equation}
\frac{8 \delta^{\mu\nu}}{-\Box + 2 A^{\mu} - 4 y^2 - 2d}
\label{propvecsing}
\end{equation}
where $a,b$ are the Dirac vector indices
\item
a Dirac singlet -- flavour vector with propagator
\begin{equation}
\frac{-8 \delta_{IJ}}{-\Box - 12 y^2 - 2d}
\label{propsingvec}
\end{equation}
where $I,J$ are the flavour vector indices.
\end{itemize}

These have the form of massive bosonic propagators, up to the
following caveat 
(of course, higher order corrections in $1/d$ may induce shifts
in the precise location of the poles, as well as their residues).

The propagators in (\ref{propvecsing}) contain the matrix $2A^{\mu}$ in the
denominator.
However, this term must be ignored since it is sub-dominant in $1/d$, compared
with the (lattice) d'Alembertian $\Box$.

The numerator of the propagator (\ref{propvecsing})
carries a delta function only, instead of the usual tensor structure
$\delta_{\mu\nu} - \partial_\mu \partial_\nu / m^2$.
This is also an artifact of the $1/d$ approximation.

Notice also that the terms which would play the role of a mass squared
in the denominators have an apparently wrong sign.
However, it is easy to check that the composite field $\epsilon_x\psi_x\psi_x$
(where $\epsilon_x = (-1)^{\sum_\mu x_\mu}$ as usual)
does lead to a massive Dirac singlet -- flavour vector propagator with
mass squared $m^2_{(0,1)} = 12 y^2 -2d = 12y^2 - 6$.
Similarly,  one obtains a massive Dirac vector --
flavour singlet with a mass squared $m^2_{(1,0)} = 4 y^2 - 2d = 4 y^2 - 6$.
We thus conclude that the right interpolating field is 
$\epsilon_x\psi_x\psi_x$~\cite{STEPHANOV}. 

The conclusion is that we find massive bound states of fermions in the
PMS phase.  They are bound by the strong interactions with the spin waves.
These composites are lighter than the elementary fermions in this phase, 
when $y$ moves away from the value $\infty$.  

\section{Conclusions}

\label{CONC}

In this work we have concerned ourselves with the general features 
and the analytical and numerical study of the lattice model given by 
expression (\ref{ACCIONZ}).

From the numerical side, we have developed a new method that exactly 
solves the technical problem related to the length-1 constraint on the 
spin variable.

The model describes qualitatively some of the properties of the doped
copper oxide compounds \cite{LETTER, HEPLAT} and has interesting properties 
in the strong coupling regime. In fact, at the Mean-Field level, no light 
fermion excitations have been identified in the FM(S) and PMS phases. 
However, in the AFM(S) phase, see section \ref{AFMSsection}, light 
excitations around momenta
 $\left(\pm\frac{\pi}{2},\pm\frac{\pi}{2},\pm\frac{\pi}{2}\right)$ have 
been found. Its possible relevance for the doped copper oxide compounds 
has been noticed in \cite{LETTER, HEPLAT,CHUBUKOV}.

Concerning the PMS phase, (see figure~\ref{PHASES}) the situation is also
 interesting. While the fermionic excitations in this phase are very heavy 
(see section~\ref{sectPMSexc}) we have found light bound states of fermions
(see section~\ref{excPMSbos}). 
They are spin singlet bosonic states of charged fermions bound by the strong 
interactions with the spin waves. A similar result has been found in the 
model of reference \cite{STEPHANOV}. 

The next step is to study the model in the presence of chemical potential and at finite temperature (after going to 3+1 dimensions). In fact, as we have proved in section~\ref{TM}, the fermion determinant is still real after the introduction of the chemical potential.

\section*{Acknowledgments}

We are indebted to L.A.~Fern\'andez for very helpful
remarks and stimulating discussions.  We also acknowledge interesting
discussions with A.~Cruz, Ph.~de Forcrand, M.A.~Stephanov and A.~Taranc\'on.
We thank, the RTNN collaboration for computing facilities.  
This work is financially supported by CICYT (Spain),
projects AEN 96-1670, AEN 96-1674, AEN 97-1680 and by Acci\'on
Integrada Hispano-Francesa HF1996-0022.


\hfill

\end{document}